\documentclass[11pt]{article}
\usepackage[normalem]{ulem}
\pdfoutput=1 

\usepackage{jheppub} 
\usepackage{url}
\usepackage{caption}
\usepackage{subcaption}
\usepackage{extarrows}

\usepackage[latin9]{inputenc}
\usepackage{float}
\usepackage{amsmath}
\usepackage{amssymb}
\usepackage{graphicx}
\usepackage{esint}
\usepackage{hyperref}
\usepackage{comment}
\usepackage{color}
\usepackage{microtype}
\usepackage{cleveref}
\usepackage{breakurl}
\usepackage{bbm}
\usepackage{amssymb,slashed,latexsym,ifpdf,amsmath}

\newcommand{\be}{\begin{equation}}
\newcommand{\ee}{\end{equation}}
\newcommand{\ben}{\begin{displaymath}}
\newcommand{\een}{\end{displaymath}}
\newcommand{\bea}{\begin{eqnarray}}
\newcommand{\eea}{\end{eqnarray}}
\def\K{K{\"a}hler }
   \newcommand{\rf}[1]{(\ref{#1})}

\newcommand{\ft}[2]{{\frac{#1}{#2}}}

\newcommand{\bbox}{\lower.2ex\hbox{$\Box$}}

\def\be{\begin{equation}}
\def\ee{\end{equation}}
\def\bea{\begin{eqnarray}}
\def\eea{\end{eqnarray}}
\def\ba{\begin{array}}
\def\ea{\end{array}}
\def\bit{\begin{itemize}}
\def\eit{\end{itemize}}

\newcommand{\lp}{\left(}
\newcommand{\rp}{\right)}
\newcommand{\ls}{\left[}
\newcommand{\rs}{\right]}

\def\rmi{{\rm i}}

 \makeatletter

\allowdisplaybreaks

\makeatother

\makeatletter
\DeclareRobustCommand{\rcite}[1]{%
  \rcite@aux#1,\@nil{#1}%
}
\def\rcite@aux#1,#2\@nil#3{%
  \if\relax#2\relax
    Ref.~\cite{#3}%
  \else
    Refs.~\cite{#3}%
  \fi
}
\makeatother

\hypersetup{
    colorlinks = true,
    citecolor = {blue},
    linkcolor = {blue},
    urlcolor = {blue},
}

 \title{\rm { \huge \bf   Goldstino Condensation? }}
 
\author[a]{Renata Kallosh,}
\author[a]{Andrei Linde,}
\author[b]{Timm Wrase,}
\author[c]{and Yusuke Yamada}

\affiliation[a]{Stanford Institute for Theoretical Physics and Department of Physics,\\ Stanford University, Stanford, CA 94305, USA}
\affiliation[b]{Department of Physics, Lehigh University, 16 Memorial Drive East, Bethlehem, PA 18018, USA
}
\affiliation[c]{Waseda Institute for Advanced Study, Waseda University, 1-21-1 Nishi Waseda, Shinjuku, Tokyo 169-0051, Japan}
\emailAdd{kallosh@stanford.edu}
\emailAdd{alinde@stanford.edu}
\emailAdd{timm.wrase@lehigh.edu}
\emailAdd{y-yamada@aoni.waseda.jp}
\abstract{ 

It was argued in \cite{DallAgata:2022abm} that the Volkov-Akulov (VA) model as well as similar models in supergravity and the related KKLT model in string theory, suffer from tachyonic instabilities due to goldstino condensation. The authors  of \cite{DallAgata:2022abm}  constructed a specific model with two unconstrained interacting chiral superfields  with linearly realized supersymmetry which has an  unstable vacuum. They  claimed that this model becomes equivalent to the VA model in the UV limit. We show that the UV  limit of their model is discontinuous, and the vacuum instability of the model proposed in  \cite{DallAgata:2022abm}  is not relevant to the VA model, to related models in supergravity, and  to the KKLT construction.}

\begin{document}

\maketitle


\section{Introduction}

The investigation of the non-linear realization of supersymmetry plays an important role in particle physics. The relation between models with linearly realized supersymmetry and the  Volkov-Akulov (VA) model \cite{Volkov:1973ix} with a global non-linearly realized supersymmetry,  was studied in great detail over the last 50 years, for example in  \cite{Rocek:1978nb,Ivanov:1978mx,Casalbuoni:1988xh,Komargodski:2009rz,Cribiori:2017ngp} and  in  Appendix \ref{appA} here. 

A consistent  supergravity with local non-linearly realized supersymmetry involving a nilpotent field was constructed only relatively recently in \cite{Bergshoeff:2015tra,Hasegawa:2015bza} and it was called  de Sitter supergravity.  An action depending on vierbein, gravitino and goldstino was constructed, and was shown to be invariant under non-linear local supersymmetry, see Appendix \ref{appB} here. 
This theory was also  derived from a superconformal model underlying supergravity with linearly realized supersymmetry \cite{Ferrara:2016een}. Lagrange multiplier superfields were introduced there in addition to physical superfields: 
once the equations of motion for the Lagrange multiplier superfields were solved, the physical superfields become constrained. The linear supersymmetry of the original models becomes non-linearly realized and its exact form was deduced from the original linear supersymmetry.

One more way to derive the de Sitter supergravity was presented in \cite{Kallosh:2015pho} using non-Gaussian integration of an auxiliary field $F$, by adding corrections to the \K potential that are of the type employed in the globally supersymmetric case in \cite{Komargodski:2009rz}. These developments are extensively used in many recent investigations related to string theory and cosmology.

A different approach to theories with non-linear realization of supersymmetry was recently proposed by Dall'Agata {\it et al}   \cite{DallAgata:2022abm}. The authors of \cite{DallAgata:2022abm} made their own attempt to derive the theory of a nilpotent superfield starting from a theory with linearly realized supersymmetry and with two unconstrained superfields. Their model suffers from a vacuum instability due to a tachyonic mass squared for the scalars. This result was interpreted in \cite{DallAgata:2022abm} as goldstino condensation and as a vacuum instability in theories with non-linear realization of supersymmetry.

The main goal of our paper is to study the relation between the models in \cite{DallAgata:2022abm} and models with non-linear realization of supersymmetry developed in~\cite{Volkov:1973ix,Bergshoeff:2015tra,Hasegawa:2015bza}. We will find that there is a discontinuity, a gap between the models with linear realization of supersymmetry proposed in~\cite{DallAgata:2022abm}, and models with the non-linear supersymmetry. Therefore, the vacuum instability found in~\cite{DallAgata:2022abm} demonstrates the problem of their specific model, but does not imply the existence of a similar instability in theories with non-linear realization of supersymmetry.

To explain it in a more detailed way, let us remember that the Volkov-Akulov model~\cite{Volkov:1973ix}, as well as the  de Sitter supergravity model \cite{Bergshoeff:2015tra,Hasegawa:2015bza},  involves a single chiral nilpotent superfield $X$ such that $X^2=0$, and supersymmetry is {\it non-linearly} realized. This theory, when formulated with a Lagrange multiplier superfield $T$, has a term in the superpotential of the form $TX^2$ as suggested in \cite{Komargodski:2009rz}.
The equation of motion for this Lagrange multiplier is $X^2=0$.
Meanwhile in~\cite{DallAgata:2022abm} the authors made an unconventional step of promoting the  Lagrange multiplier superfield  $T$  to the status of a normal propagating superfield: such a theory has two superfields $X$ and $T$ with {\it linearly} realized supersymmetry. 
  As a result of this and several other unconventional modifications made in~\cite{DallAgata:2022abm}, the  models they proposed suffer  from a vacuum instability. 
  
  This result by itself does not say anything about a vacuum stability in models with non-linearly realized supersymmetry \cite{Volkov:1973ix,Bergshoeff:2015tra,Hasegawa:2015bza}. To extend their results to the models  \cite{Volkov:1973ix,Bergshoeff:2015tra,Hasegawa:2015bza} the authors of \cite{DallAgata:2022abm} are using   the so-called  exact renormalization group (ERG) equations, following the rarely used procedure proposed in \cite{Jaeckel:2002rm}.
    
  Note that the  goldstino condensation proposal made in \cite{DallAgata:2022abm} for the VA theory has important features distinguishing it from the Nambu-Jona-Lasinio (NJL) type models \cite{Nambu:1961tp, Nambu:1961fr} used more commonly for studies of  condensates and bound states. In models where a large $N$ expansion is available, one usually adds auxiliary fields  which allow a bosonization of the fermionic models. Such auxiliary fields have terms quadratic and linear in the action, they do not have kinetic terms. The next step is a 1-loop computation of the diagrams involving the original fields coupled to an auxiliary field. This computation produces a kinetic term  for an auxiliary field, as well as  other terms in the effective action. See for example \cite{Coleman:1974jh,Kobayashi:1975ev,Abbott:1975bn,Eguchi:1976iz,Linde:1976qh,Kirzhnits:1978fy,Bardeen:1989ds,Ebert:1994sm} where the bosonic and fermionic bound states were investigated in the large $N$ approximation, which allowed to avoid significant problems with using the renormalization group approach in this context.

Since VA theory has only one fermion, the authors of  \cite{DallAgata:2022abm}   could not use the large $N$ approximation to support their statements.  
That is why  they  refer to \cite{Jaeckel:2002rm} as the only case in the literature where  the NJL type model  was discussed for a single fermion $N=1$ in a non-perturbative renormalization group flow. 
 And they argued that the  ERG  equations in the UV limit may continuously relate the unstable models they proposed in  \cite{DallAgata:2022abm}  to the models with non-linearly realized supersymmetry~\cite{Volkov:1973ix,Bergshoeff:2015tra,Hasegawa:2015bza}.

It is not our goal to debate the reliability of the renormalization group equations in the context of a non-renormalizable theory, though we have some concerns about its consistency,  to be discussed later. In the main part of  this paper we will simply follow the lead of \cite{DallAgata:2022abm} and confirm many of their results obtained in the context of the models with two unconstrained interacting chiral superfields $X$ and $T$. However, we will also show that these models with linearly realized symmetry and tachyons do not coincide with the models with the non-linearly realized supersymmetry \cite{Volkov:1973ix,Bergshoeff:2015tra,Hasegawa:2015bza} in the UV limit. This discontinuity invalidates the main conclusions of \cite{DallAgata:2022abm}.

To understand the origin of the discontinuity, which we will discuss in great detail in this paper,  consider a simple toy model with the following \K potential $K$ and superpotential $W$:
\be
K=  X  {\bar X} +  t^{2} T  {\bar T},  \qquad \qquad W   =   f X.
\label{GD1} \ee
Here $t$ and $f$ are some  parameters.

One could expect,  that in the limit $t\to 0$ the term $ t^{2} T  {\bar T}$   disappears, and the model becomes a theory of a single field $X$ with
\be
K=  X  {\bar X} ,  \qquad \qquad W   =   f X \ . 
\label{GD2} \ee
However, such conclusion would be premature.  
Taking the limit $t\to 0$ in the \K potential \rf{GD1} 
is problematic 
since  the \K metric tends to zero  in this limit,  $K_{T\bar T} \to 0$,  and the inverse one is divergent, $K^{T\bar T} \to \infty$.

The field $T$ in \rf{GD1} is not canonically normalized. Switching to canonical variables for $T$ in \rf{GD1} by making a field redefinition $t T \to T$, which is a valid procedure for any {\it finite} $t$, one finds that the model \rf{GD1} for any  $t\not = 0$ is equivalent to the model  
\be
K=  X  {\bar X} +  T  {\bar T},  \qquad \qquad W   =  fX .
\label{GD3} \ee
This model describes two canonically normalized non-interacting massless fields $X$ and $T$.

Thus we see that the theory \rf{GD1}  for $t = 0$, given in  \rf{GD2}, is {\it not} equivalent to the theory \rf{GD1} in the  limit $ t\to 0$ given in \rf{GD3}. In other words, the limit $ t\to 0$ in the model \rf{GD1} is discontinuous.  

The existence of this discontinuity could suggest that if one adds  the term $ t^{2} T  {\bar T}$ to the models such as \rf{GD2}, the field $T$ is there to stay, and one cannot get rid of it in the limit $t\to 0$. This rule is generally valid, but there are some possible exceptions. As an example, consider  a theory 
\be
K=  X  {\bar X} +  t^{2} T  {\bar T},  \qquad \qquad W   =   f X + T^{2}.
\label{GD4} 
\ee
If one takes $t= 0$, the field $T$ no longer propagates, its equation of motion becomes $T=0$, and the theory reduces to \rf{GD2}.

On the other hand, for any finite $t$ this theory is equivalent to
\be
K=  X  {\bar X} +  T  {\bar T},  \qquad \qquad W   =   f X + {T^{2}\over t^{2} } \qquad \Rightarrow \qquad V= f^2 + 4 {T\bar T\over t^4}\ .
\label{GD5} 
\ee
In the limit $t\to 0$ this theory describes a massless field $X$ and a massive field $T$ with the mass $m_{T}$ proportional to $1/t^{2}$, and with the potential having a minimum at $T = 0$. In the limit $t\to 0$ this field becomes infinitely heavy  and decouples from all dynamical processes. In that  sense,  the model \rf{GD4} in the limit $t\to 0$ may become  effectively equivalent to the model \rf{GD2}. However, even in this case it is hard to reliably establish  the equivalence of the two models  because quantum corrections to the potential $V$ typically blow up in the limit $m_{T} \to \infty$.

These examples will play a crucial role in our analysis of the investigation performed in \cite{DallAgata:2022abm}. Indeed, the authors considered the Volkov-Akulov model \cite{Volkov:1973ix} in a form proposed  in \cite{Komargodski:2009rz}
\be
K=  X  {\bar X} , \qquad W=  fX + {1\over 2} TX^2\ ,
\label{VA1} \ee
where the field $T$ is not a physical degree of freedom but a Lagrange multiplier.
Then they  modified the  first term of the \K potential in eq. \rf{VA1} and 
added to it several new terms which were absent in the original VA theory: 
\be
K= \alpha(t)  X  {\bar X} +   \beta(t)  T  {\bar T} + g(t)  X\bar X T\bar T + {q(t)\over 4}( X\bar X)^2\ , \qquad W=  fX + {1\over 2} TX^2\  ,
\label{GD} \ee
with   $\alpha(t) >1 ,\  \beta(t) > 0, \  g(t) > 0,  \   q(t) >0$. 

In this new model the field $T$ is no longer a Lagrange multiplier, it becomes a new propagating field. This modification is a   substantial and highly nontrivial step, as we illustrated by the toy models discussed above. One should not take for granted that the new theory is equivalent to the original one even when the limit  
\be
 \alpha(t) \to 1\, , \qquad \beta(t) \to 0   \, , \qquad g(t) \to 0  \,   , \qquad q(t) \to 0\,  , \qquad  {\rm at } \, \, \, t  \to 0
\label{UV}
\ee
 is taken  in the \K potential.  Meanwhile that is exactly what the authors of \cite{DallAgata:2022abm} did,  and that is what they call a UV limit. 
 
The authors interpreted the parameters of their model in eq. \rf{GD} as  effective coupling constants in the RG equations with respect to the running parameter $t = \ln {\Lambda\over \mu}$, where $\mu$ is the normalization mass scale, and $\Lambda$ is  a cut-off. Then they found  specific solutions of  the RG equations such that  eq. \rf{UV} is valid. On the basis of this result  they conjectured that the model \rf{GD}  coincides  with the VA theory in the UV limit. And since they also found that their own theory \rf{GD} suffers from the tachyonic instability at any $t$ at $X=T=0$, they  conjectured that the VA  model has a similar problem.

However, in the UV limit  a component of a  \K geometry vanishes  and its inverse is singular:
\be  
 g_{T\bar T}=K_{T\bar T} = \beta(t) + g(t) X\bar X\to 0\, ,  \quad g^{T\bar T}\to \infty  \quad  {\rm at } \, \, \, t  \to 0 \ .
\label{sing}\ee
But supersymmetry and supergravity theories are well defined only for a non-singular \K geometry. Therefore the model \rf{GD} is not well defined at $t \to 0$. This is yet another manifestation of the issue illustrated by our toy models  \rf{GD1}-\rf{GD5}, which demonstrated that adding new terms such as   $\beta(t)  T  {\bar T}$ to the \K\ potential may  irreversibly alter the physical content of the model, and make a transition from the VA model \rf{VA1} to the model \rf{GD} discontinuous. 

To further explore the model \rf{GD} at small (but finite) $t$  one may  switch to canonical variables at $t\neq 0$, as we did in the toy example in eqs. \rf{GD4}, \rf{GD5}, and study this model and  its properties. No such analysis of the possible discontinuity between the models \rf{VA1} and \rf{GD}, or between their supergravity generalizations,  was performed in   \cite{DallAgata:2022abm}, and the issue of the \K metric singularity in the UV limit presented  in eq. \rf{sing} was never raised. This issue requires a new detailed investigation, which is the goal of our paper. 

The main conclusion of our investigation is that the limit $t\to 0$ is discontinuous, and therefore the vacuum instability of the model proposed in \cite{DallAgata:2022abm} is not relevant to the VA model, to related models in supergravity, and  to the KKLT construction.

We would like to add a comment here about the relation between the VA model and a single superfield model with a linearly realized supersymmetry, where $X$ is not a nilpotent superfield \cite{Komargodski:2009rz}, but the theory has $(X\bar X)^2$ correction to \K potential originating from a certain one-loop diagram,   see eq. \rf{KS} in  Appendix \ref{appA}. The model \rf{KS} coincides with   \rf{GD} with  $T=0$,  $\alpha=1$, ${q(t) \over 4} =-{c \over M^2}$.
The one-loop corrections are known to lead to  $c>0$ in the model \rf{KS},  and to positive scalar mass squared $m^2 = 4c {f^2\over M^2}$. It is widely accepted that the one-loop corrections bring this model with linear supersymmetry to the VA theory with a non-linear supersymmetry. Scalars are heavy and decoupled at large positive $c$. In Ref.  \cite{Cribiori:2017ngp}  `From Linear to Non-linear SUSY'  the constant $c$ was also taken to be positive. But in the recent paper \cite{DallAgata:2022abm} two of the authors of  \cite{Cribiori:2017ngp} decided to take  $c< 0$ for the consistency of their ERG equations in the presence of the superfield $T$. This change of sign of a quartic coupling in the \K potential is not supported by known quantum corrections in these models, see Appendix \ref{appA} for details.  After this change of the sign of $c$, the scalars in a one-superfield model with $T=0$  become tachyonic, and the model is no longer related to the VA theory.

Independently of the above, we also point out that the results obtained in \cite{DallAgata:2022abm} cannot be extrapolated to the anti-D3-brane uplift in the KKLT construction \cite{Kachru:2003aw}. The reason is the following shortcoming stated at the end of the introduction in \cite{DallAgata:2022abm}: light states surviving in the IR can disrupt their procedure. Given that the SUSY breaking scale for the anti-D3-brane in the KKLT scenario is the warped down string scale, which is above the warped down KK scale, there are plenty of light fields below the SUSY breaking scale. In particular, the world volume fields on the anti-D3-brane contain in addition to the goldstino a massless U(1) gauge field. Such light fields have been neglected in the analysis of \cite{DallAgata:2022abm}. We will discuss below in section \ref{sec:string} explicit examples where the presence of light states ensures the absence of potential instabilities and we argue that this should also apply to the anti-D3-brane in KKLT.

\section{Lagrange multiplier and instability}\label{sec: LM}

Let us consider a theory with a Lagrange multiplier $T$ in  eq. \rf{VA1}. It was given in this form in the globally supersymmetric case in \cite{Komargodski:2009rz}. A Lagrange multiplier is a field which appears linearly in the action. Therefore, it is different from auxiliary fields which appear quadratically. The equation of motion for $T$ means that the factor in front of $T$ has to vanish, which is a nilpotency equation for the superfield $X$
\be
X^2=0 \ ,
\ee
and VA theory is restored. Note that  $T$ is not uniquely defined, which is also different from the situation with the auxiliary field. 

Following \cite{DallAgata:2022abm} we add to this theory a kinetic term for the Lagrange multiplier $T$  
\be
K= X\bar X + Z_T \, T\bar T\ ,  \qquad W= fX + {1\over 2}TX^2 \ .
\label{Yu} \ee
Now,  if $Z_T \neq 0$ as suggested in \cite{DallAgata:2022abm} we cannot solve the equation for $T$ as it is not algebraic anymore. We have a model with two unconstrained coupled superfields $X, T$. This is a simplified version of the model studied in \cite{DallAgata:2022abm} in their eqs. (3.1), (3.2), where we can take $ \tilde \zeta=0, \tilde \gamma=0$ to get from eq. \rf{Yu}
\be
K=  X {\bar X} +    T  {\bar T}  \ , \qquad 
W= f  X + {1\over 2 \sqrt {Z_T} }  T  X^2 \ .
\label{SuS}\ee
The model in  \rf{SuS} at any $Z_T \neq 0$ is equivalent to the model in \rf{Yu}.
The  term added to the \K potential of the VA theory in \cite{DallAgata:2022abm} is of the form $Z_T T \bar T$. To make it canonical one had to rescale it so that 
$Z_T T \bar T \to T\bar T$. Therefore the couplings of the form $TX^2$ was rescaled as $TX^2 \to {1\over Z_T^{1/2} } TX^2$. The vertex involving the $T$-scalar and 2 fermions of the $X$ superfield and a vertex involving the scalar from the $X$ multiplet with fermions from the $T$ and $X$ multiplets blow up in the limit $Z_T \rightarrow 0$. 
Already in this simple model we  see the issues with the claim in   \cite{DallAgata:2022abm} that the model where $T$ is a propagating field is equivalent to a model where $T$ is a Lagrange multiplier.

In the supergravity version of this model with $K, W$ in eq. \rf{Yu} the masses squared of the 4 canonical real scalar fields at $X=T=0$  (with double multiplicity) are blowing up in the limit $Z_T \to 0$
\be
m^2_{\pm} ={1\over 2} f 
\left. \lp f  \pm \sqrt{f^2 +{ 4\over Z_T } }\,\rp \right |_{Z_T \rightarrow 0}  \, \rightarrow \,  \pm \,  {f \over \sqrt {Z_T}} \ .
\label{mpm}\ee
There are 2 tachyons with negative masses squared at fixed values of $Z_T$. In the limit $Z_T \rightarrow 0$ all masses diverge: this means that 2 states with positive masses squared decouple, but two states with negative masses squared grow exponentially.

The heavy scalars in equation \rf{mpm} have the time-dependent wave functions proportional to \footnote{We use the notation from \cite{DallAgata:2022abm} where $t_{RG}\equiv t= \log {\Lambda\over \mu}$ everywhere except in eqs. \rf{time}-\rf{exp} below, where $t$ denotes the normal time. }
\be
e^{i \sqrt {\vec k^2 +m^2_{\pm} } \, \, t} \ .
\label{time}\ee
At large positive $m^2_+$ these are superheavy particles. In the large mass limit they are expected to immediately decay, and it is hard to produce them, even as virtual particles. 
 
On the other hand, tachyons are particles that have negative mass squared $m^2_-$. Quantum fluctuations of the tachyonic states with $|m^2_-|> \vec k^2$ have components growing as 
\be
e^{\sqrt {|m^2_-| - \vec k^2} \, \, t}  \ .
\label{time1}
\ee
At small $\vec k$  these modes grow exponentially as 
\be
 e^{ \mathcal{M} \, t}\, ,  \qquad  \mathcal{M} \equiv  |m^2_-|^{1/2} \ .
\label{exp}\ee
In our case in equation \rf{mpm} we have $\mathcal{M}^2={f\over \sqrt{Z_T}}$. An increase of the tachyonic mass when $Z_T \rightarrow 0$ does not lead to decoupling. Instead of that, quantum fluctuations of the tachyonic scalars grow exponentially, as $e^{ \mathcal{M} \, t}$, which leads to immediate decay of the initial vacuum state  within the time $O(1/ \mathcal{M} )$ \cite{Felder:2001kt}.

In the global case we find the masses squared of the scalar fields at vanishing scalar vevs are (with double multiplicity) 
\be
m^2_{\pm} = \pm  {f \over \sqrt{Z_T}} \ ,
\ee
confirming the fact that, instead of decoupling, two of the four real scalars grow exponentially in the limit to the VA theory.

\section{The model proposed in  \cite{DallAgata:2022abm}}\label{sec4}
The complete model in \cite{DallAgata:2022abm} where the \K potential is given in their eq.  (3.1) and the superpotential in their eq. (3.2) is
\be
K=  X {\bar X} +    T  {\bar T} + \tilde \gamma T  {\bar T}  X  {\bar X}+  \tilde \zeta  X {\bar X}  X  {\bar X} \ ,
\label{Ka}\ee
\be
W=\tilde f  X + \tilde g  T  X^2 \ .
\label{Su}\ee
We present in Appendix \ref{appC} the steps in the derivation of this model according to \cite{DallAgata:2022abm} starting with the VA theory where the term $T\bar T $ is absent in the \K potential and $\tilde \gamma= \tilde \zeta =\tilde g -{1\over 2} =0$. It is first added in the form $Z_TT\bar T$ and $Z_T$ is rescaled away afterwards.\footnote{Here we do not use the $\hat T$ notation as in eq. \rf{S2Far0}, for simplicity.} The model with the added kinetic term for the Lagrange multiplier is the starting point for the ERG approach.

We note here that there is only one point in the moduli space geometry where the \K geometry at $t\to 0$ is not singular. Namely only at the the point $T=X=0$ the \K geometry is regular in what is called UV limit in  \cite{DallAgata:2022abm}. We will see below that at any  non-vanishing vev of the scalars the \K metric in the UV limit in  \cite{DallAgata:2022abm} is singular and the models are not well defined in the UV limit.

The coupling constants in this model are function of the renormalization group time 
\be
t_{RG}\equiv t= \log {\Lambda\over \mu}\geq 0 \ ,
\ee 
and 
\be
{f\over \Lambda^2} \equiv \xi_{UV}\geq 1\ .
\ee

Here the theory has  a cut-off  $\Lambda$ and $\mu\leq \Lambda $ is the  renormalization scale,  related to the energy scale at which  the theory is probed.
As a result of solving the ERG equations the model has the following $t$-dependent couplings:
\be
\tilde \zeta =\frac{1-e^{-2 t}}{4 \left(\frac{t+\frac{e^{-2 t}}{2}}{8 \pi ^2}-\frac{1}{16 \pi ^2}+1\right)^2}\, ,  \qquad 
\tilde \gamma =\frac{1-e^{-2 t}}{\left(\frac{t+\frac{e^{-2 t}}{2}}{8 \pi ^2}-\frac{1}{16 \pi ^2}+1\right) \left(\frac{t+\frac{e^{-2 t}}{2}}{16 \pi ^2}-\frac{1}{32 \pi ^2}\right)}  \ ,
\label{tdep1}\ee

  \be
\tilde f=\frac{f  e^{2 t}}{\Lambda^2 \sqrt{\frac{t+\frac{e^{-2 t}}{2}}{8 \pi ^2}-\frac{1}{16 \pi ^2}+1}} \, , \qquad 
\tilde g=\frac{1}{2 \left(\frac{t+\frac{e^{-2 t}}{2}}{8 \pi ^2}-\frac{1}{16 \pi ^2}+1\right) \sqrt{\frac{t+\frac{e^{-2 t}}{2}}{16 \pi ^2}-\frac{1}{32 \pi ^2}}} \ .
\label{tdep2}\ee
The  renormalization group solutions proposed in the paper  \cite{DallAgata:2022abm} require that
 \bea
\tilde \gamma  >0\, , \quad
\tilde \zeta  >0\, , \quad
\tilde f  >0\, , \quad
 \tilde g  >0\, .
\label{pos}\eea
We will focus on the limit $t\to0$ where the renormalization scale $\mu$ is infinitesimally away from the cut-off $\Lambda$. As we will show below, the singular behavior of the scalar potential becomes more significant as $\mu$ approaches the cut-off scale $\Lambda$. This implies a discontinuity between the VA theory at $Z_T=0$, i.e. at $t=0$, and the model of \cite{DallAgata:2022abm} for $Z_T \to 0$, i.e. for $t\to0$. In the limit of small $t$ we find from \rf{tdep1}, \rf{tdep2} that
\be
\tilde \zeta  \to {t\over 2}\ , \qquad 
\tilde \gamma  \to {32 \pi^2\over   t}\ , \qquad 
\tilde f \to {f \over \Lambda^2 } \ , \qquad 
\tilde g \to {2\pi\over   t} \ .
\label{smallt}\ee
The couplings of  $T$ and $X$, namely the couplings  $\tilde \gamma, \tilde g$  tend to infinity in the limit $t\rightarrow 0$. At the qualitative level  this is understandable since the original term added to the \K potential of the VA theory was of the form $Z_T T \bar T$. To make it canonical one had to rescale it so that 
$Z_T T \bar T \to T\bar T$. Therefore the couplings of the form $TX^2$ was rescaled as $TX^2 \to {1\over Z_T^{1/2} } TX^2$ and $T\bar T X\bar X \to  {1\over Z_T } T\bar T X\bar X$. More details are given in Appendix~\ref{appC}.

The supergravity  model in \cite{DallAgata:2022abm} with two unconstrained superfields has the  \K potential and superpotential given in  eq. \rf{Ka}, \rf{Su}.  The potential, according to  \cite{DallAgata:2022abm} has the standard form,  the only difference is the value of $\tilde M_{Pl}$
\be
V= e^{K\over \tilde M_{Pl}^2} \left( DW_i g^{i{\bar \jmath} } {\overline {DW}}_{\bar \jmath}  - 3 {|W|^2\over \tilde M_{Pl}^2}\right) \ .
\label{pot}\ee
They propose to use $\tilde{M}_{Pl}=\Lambda e^{t} P$ where $P$ is some `realistic value' dimensionless parameter, for instance, $P \simeq 10^4$. This is not relevant for the computation of the mass spectrum since both the normal $M_{Pl}^2$ as well as $\tilde {M}_{Pl}^2$ are space-time independent. 

The supergravity version of the mass formula was not given in \cite{DallAgata:2022abm}, but it was observed that as in the rigid case, the masses are highly tachyonic. We present here the masses of the canonical scalar fields, in units $\tilde M_{Pl}=1$, at $X=T=0$. Here $\tilde M_{Pl}= e^t P$ and we are interested in values of $t$ close to zero where the dependence of  $\tilde M_{Pl}$ on $t$ can be ignored.

There are 4 real scalars, the masses squared are doubly degenerate and given by
\be
m^2_{\pm}={1\over 2} \tilde f ^2\left[ A \pm \sqrt {A^2 + B}\right] \ ,
\ee
where
\be
A= 1 -\tilde \gamma  -4 \tilde \zeta  \ ,
\ee
\be
B=16 \left(  {\tilde g^2\over \tilde f^2} + (1- \tilde \gamma )   \tilde \zeta\right)\, , \qquad A^2+B= 16   {\tilde g^2\over \tilde f^2} + (1- \tilde \gamma +   4\tilde \zeta)^2 \ .
\ee
We confirm,  based on eq. \rf{pos}   that  there are tachyons in the supergravity model in \cite{DallAgata:2022abm} at $X= T=0$. 

However, if we would consider the supergravity model  in  eqs. \rf{Ka}, \rf{Su} without assuming the choice of parameters made in \cite{DallAgata:2022abm} based on the ERG approach, we would observe that for 
\be
 1- \tilde \gamma  > 4\tilde \zeta  \quad       \Rightarrow  \quad   A>0, \qquad 
-( 1-   \tilde \gamma    ) \tilde \zeta > {\tilde g^2\over \tilde f^2}  \quad      \Rightarrow  \quad       B<0 \ ,
\label{2}\ee 
all 4 masses squared are positive and there are no tachyons. Both conditions are violated at small $t$ as we can see from eq. \rf{smallt}.

At small $t$ we find 2 heavy states and 2 tachyons
\be
m^2_{\pm} |_{t \,  \rightarrow  0} =  -  {  1\over  t    }  \left[ {1   \pm  \sqrt { 1+l }\over l }   \right ] \ , \qquad l \equiv {   \Lambda ^4\over  16 \pi^2 f^2 } \ .
  \label{tach1lim1}\ee
The total scalar potential in \rf{pot} at $X=T=0$ for $t\to 0$ can be computed and it is constant
\be
V|_{X=T=0}={f^2\over \Lambda^4} \ .
\ee
However, if we consider a general situation  where both of the complex scalars $X$ and $T$ do not vanish, then the leading term in the small $t$ limit is singular, blowing up as
\be
V |_{t \,  \rightarrow  0} =  {128\pi^4 \over  t^3}|X|^6\, |T|^4  e^{32\pi^2  |X|^2\, |T|^2\over t} \ .
\label{small_t}\ee

The \K geometry $g_{i\bar j} =\partial_i\partial_{\bar \jmath} K$ defining the kinetic terms for the  scalars and fermions
is also singular at $t\to 0$
 \be
g_{i\bar \jmath} = \left(
\begin{array}{cc}
1+ 32 \pi ^2  T\bar T\lp \frac1t-\frac13 \rp& \, \, \, 32 \pi ^2 T \bar X \lp \frac1t-\frac13 \rp\\
 \\
 32 \pi ^2 \bar T  X \lp \frac1t-\frac13 \rp& \, \, \, 1+ 32 \pi ^2  X\bar X\lp \frac1t-\frac13 \rp\\
\end{array}
\right) + \mathcal{O}(t)\ .
\ee
We can also compute the part of the action  that is quadratic in the fermions 
\be
{\cal L}_{ferm} = {1\over 2} m_{ij} \bar \chi^i  \chi^j + h. c. \, , 
\qquad m_{ij}\equiv e^{K\over 2} D_{i} D_{j} W \ .
\label{ferm}\ee At the saddle point with $X=T=0$ all these terms vanish, the fermions are massless. However, at small $t$ and non-vanishing scalar vevs we find 
\be
m_{ij}|_{t \,  \rightarrow  0} = {2048 \pi ^5  |X|^4  |T|^2 \bar{T}\over t^3} e^{  16 \pi ^2 |X|^2\, |T|^2 \over  t}
\lp \begin{array}{cc}
   T^2 & T X\vspace{2mm}\\\vspace{2mm}
   TX & X^2
  \end{array}\rp \ ,
\ee
and the eigenvalues of the matrix $m_{ij}$ are  
\be
{\rm Eigenvalues}[  \, m_{ij}|_{t \,  \rightarrow  0} ] = {2048 \pi ^5  |X|^4  |T|^2 \bar{T}\over t^3} e^{  16 \pi ^2 |X|^2\, |T|^2 \over  t}
\label{eig}\{  0, \, \, T^2+X^2\}\ .
\ee
One is vanishing, the other is singular.\footnote{In the field space coordinate system $(X,T)$, the eigenvectors for each eigenvalue $\propto \{0,T^2+X^2\}$ are $(-X/T,1), (T/X,1)$. Therefore, the mass eigenmodes are mixtures of $\psi^X$ and $\psi^T$ for general values of $X,T$.} 
The critical points of the potential at small $t$ are not available for non-vanishing scalar vev's. But we will find  that in the small $t$ limit, there are  flat directions and massless bosons.

It is possible to change variables to make the kinetic term canonical, non-singular and to express the rest of the action in these variables. The potential and the fermion action at small $t$ are complicated and might still involve terms singular in the $t\to 0$ limit. 
But since there is no extremum of the potential at $X\neq 0$ and $T\neq 0 $ for $t\to 0$ the procedure of switching to canonical variables does not help to define the physical states and the masses of these canonical fields. It only shows that the limit to theories with non-linear supersymmetry from the model in \cite{DallAgata:2022abm} is discontinuous: the scalars go to 0 first and $t\to 0$ afterwards, or vice versa, the results are different.

Thus we have now shown that the system at $X=T=0$, which is a saddle point of the potential, is highly unstable. This is due to the tachyons with blowing up negative masses at $t\to 0$ as we see in eq. \rf{tach1lim1}.
We have also shown that the model defined by eqs. \rf{Ka}, \rf{Su} and studied in \cite{DallAgata:2022abm}, which according to this paper represents de Sitter supergravity at $t=0$, actually does not have a well-defined limit $t \to 0$. In particular, the tachyonic masses of the canonical scalar fields blow up in this limit. Also the action at non-vanishing scalar values diverges for $t\to 0$, but at vev's $X=T=0$ the action is finite. Of course we still have quantum vertices with scalar and fermion couplings proportional to $\tilde g \sim {1\over t}$ and to $\tilde \gamma \sim {1\over t}$.

In the globally supersymmetric model the potential is  based on the same \K potential and superpotential given in eqs. \rf{Ka}, \rf{Su}
\be
 V=  g^{i\bar \jmath} \partial_i  W \partial_{\bar \jmath} { \overline W} \ .
 \ee
  At the critical point $ X=T=0$  we have  found the following double set of masses squared for the canonical scalar fields \footnote{Our masses differ from the the ones in \cite{DallAgata:2022abm} by a factor of 1/2.}   
\be
 m^2_{\pm} = - {1\over 2} \tilde f^2 \ls ( \tilde \gamma
 + 4 \tilde \zeta) \pm \sqrt { {16\tilde g^2\over \tilde f^2} + (\tilde \gamma - 4 \tilde \zeta)^2}  \rs \ .
  \label{tach1}\ee
In a model in \cite{DallAgata:2022abm} where  eq. \rf{pos} is valid, one can see that  there are tachyons in \rf{tach1}. At small $t$ the mass formula is the same as in equation \rf{tach1lim1} in the local case.

We find the same type of discontinuity in the properties of the total potential and the fermionic action of the globally supersymmetric model. Depending on the order of the limits the total potential and fermion action either are finite
at $ X=T=0$, $t \,  \rightarrow  0$ or  singular at $t \,  \rightarrow  0$ with $ X\neq 0, T\neq 0$. This means that the limit to the VA theory from the theory in \cite{DallAgata:2022abm} has a discontinuity.

\begin{figure}[H]
\centering
\includegraphics[scale=0.34]{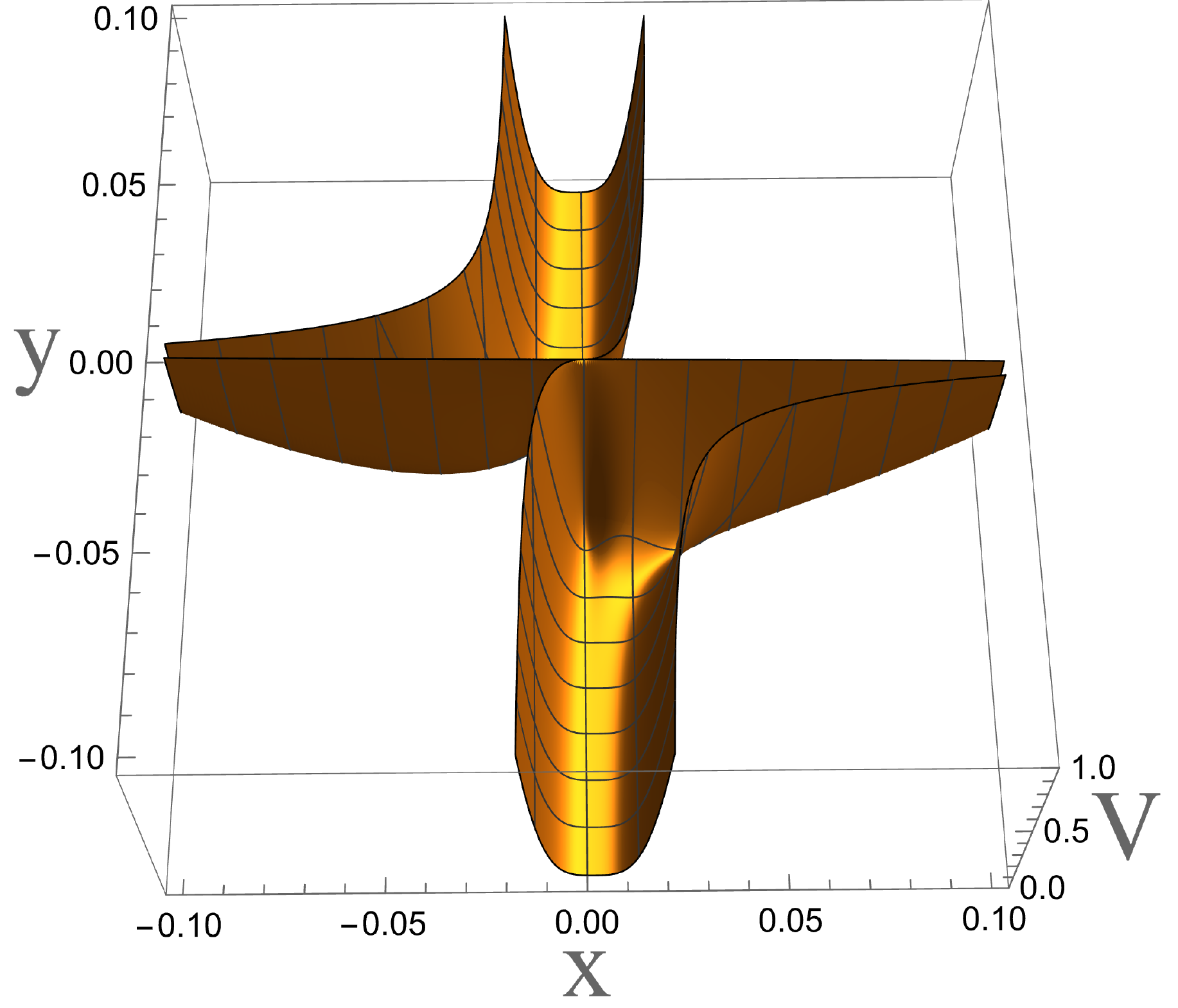} \qquad \includegraphics[scale=0.34]{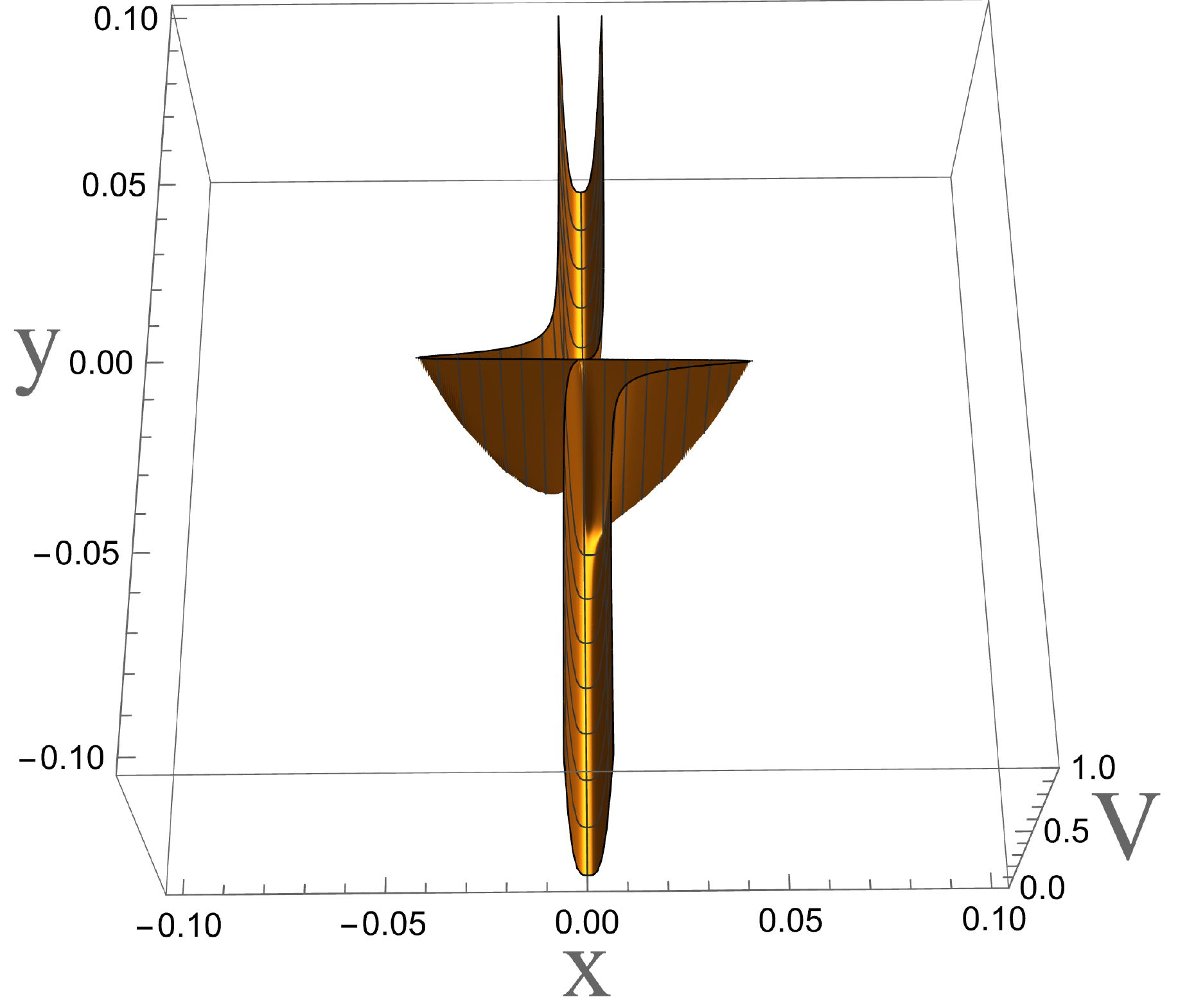} 
\caption{\footnotesize Here we plot the potential in eq. \rf{small_t} for $X=\bar X=x$, and $T=\bar T=y$.
On the left $t=0.001$, on the right $t=0.0001$}
\label{XTsmt}
\end{figure}

We studied also the case of $X=\bar X=x$, $T \neq 0$. Using the R-symmetry one can choose the vev of the $X$ field to be real. We have found that the minimum of the potential is at  $X\sim t$ for non-vanishing $T$. The corresponding potentials are shown in Fig. \ref{XTsmt} for the case $X=\bar X=x$, $T= \bar T \neq 0$ at decreasing time.
The potential at $X=\bar X=x$ as a function of $T=\bar T =y$ has a flat direction as it is proportional to $t$ at all values of $T\neq 0$. It means that at $t=0$  along the valley $V=0$  there is a massless scalar field in  \cite{DallAgata:2022abm}, which is absent in the VA theory. Once again we have shown that the model in \cite{DallAgata:2022abm} is not representing the VA theory with a non-linear supersymmetry. 

\begin{figure}[H]
\centering
\includegraphics[scale=0.63]{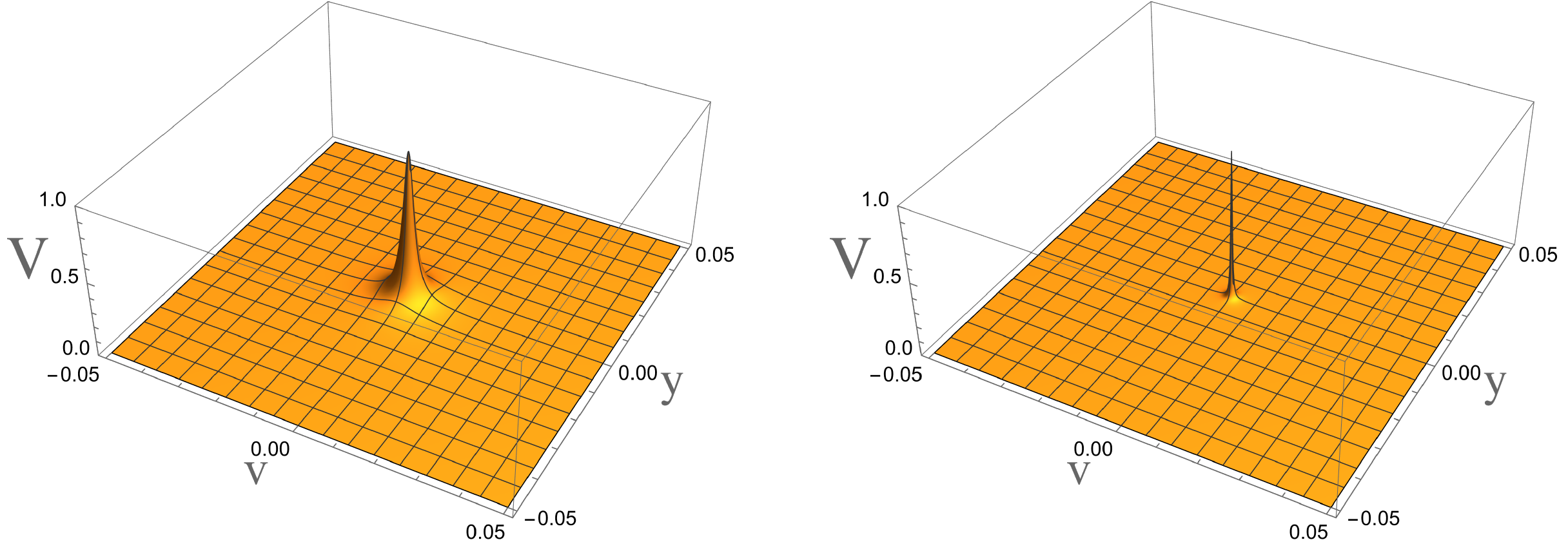} 
\caption{\footnotesize Here we plot the potential  for $X=\bar X=0$, and $T={1\over \sqrt 2} (y+iv)$.
On the left $t=0.0005$, on the right $t=0.00001$. The potential  becomes a tiny needle peaking out of a flat plane in the $t\to 0$ limit.}
\label{Tpotential}
\end{figure}

We have also looked at the potential at $X=\bar X=0$ as a function of $T={1\over \sqrt 2} (y+iv)$. The corresponding potential at different values of small $t$ is shown in Fig. \ref{Tpotential}. One can see a flat complex plane once the field $T$ rolls down from an unstable $T=0$ position. In terms of the massless scalars we find that the theory at small $t$ has 2 massless scalar degrees of freedom. This is, as expected, very different from the content of physical states in the VA theory where there are no massless scalars. 

 \section{Scale of energies and numerics}\label{sec:num}
\begin{figure}[H]
\centering
\includegraphics[scale=0.35]{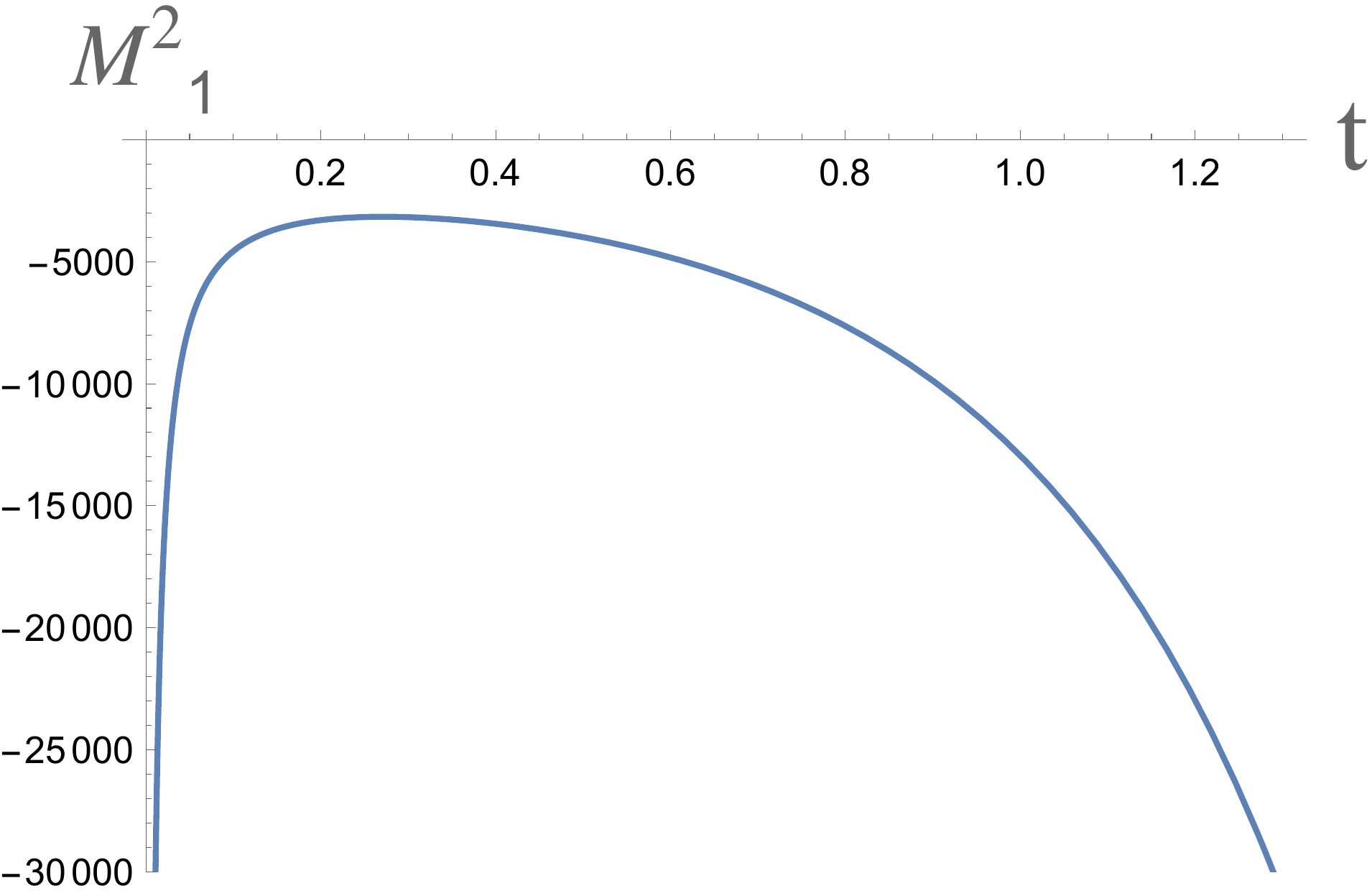}   \qquad \ \includegraphics[scale=0.35]{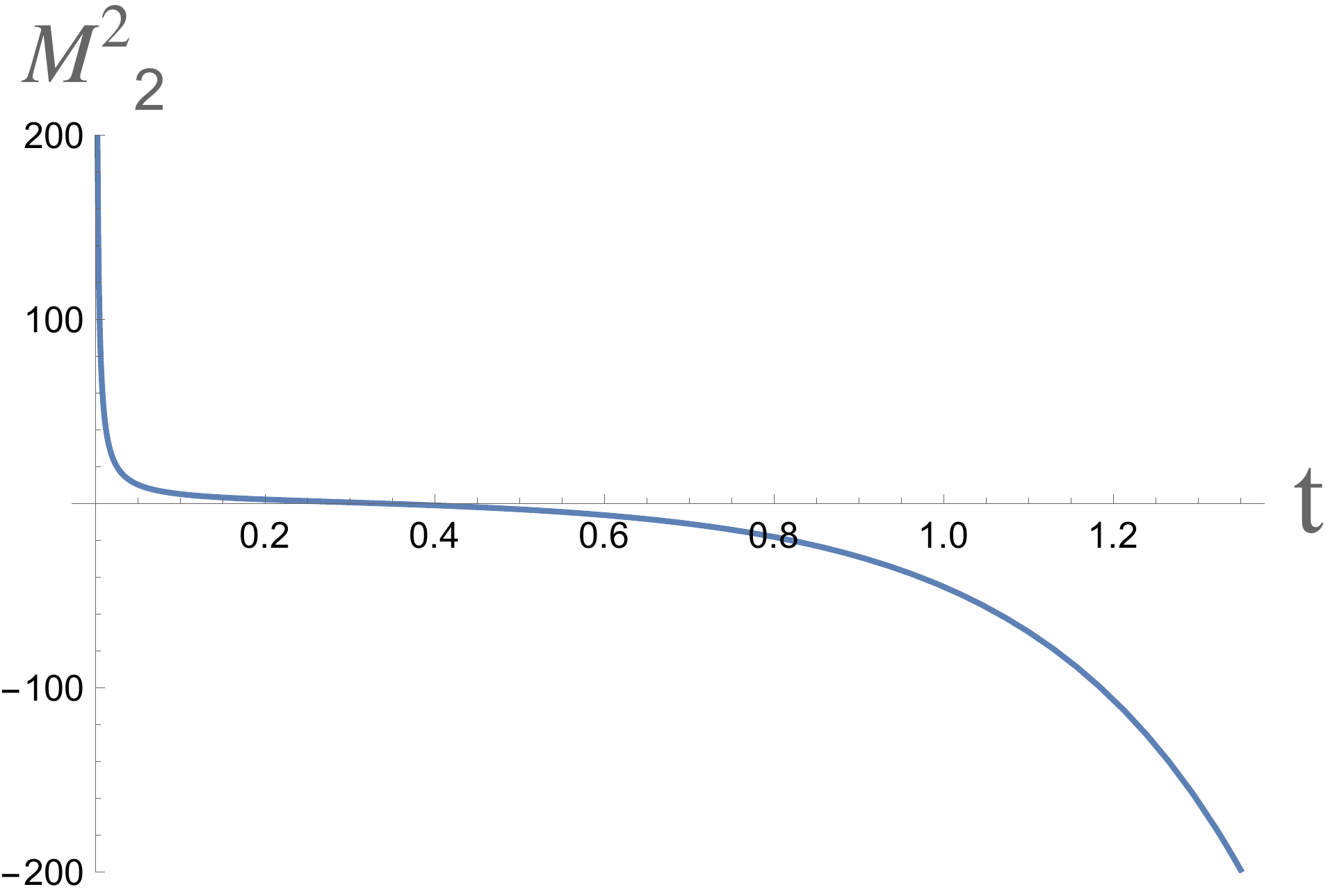} 
 \vskip  5pt
\caption{\footnotesize Here we plot the numerical values of the scalar masses as function of $t$ for $f=\Lambda^2$.}
\label{Masses}
\end{figure}

The goldstino condensate construction in \cite{DallAgata:2022abm} is based on the assumption that there is a cut-off at $\Lambda$ and that the ERG evolution is in terms of the RG time $t = {\Lambda\over \mu}$ where $t$ is non-negative and the momenta $p^2$ are restricted to be below $\mu^2$, since the theory is probed at energies below $\mu^2$
\be
\mu^2\leq \Lambda^2\ ,  \qquad p^2  < \mu^2\ .
\ee
Here we take the results for the masses of the scalar fields obtained in \cite{DallAgata:2022abm} at their face value to find out how these masses are related to the cut-off $\Lambda^2$.
We take the mass eigenvalues in eq. \rf{tach1} and use the dependence on $t$ of all entries there given in eqs. \rf{tdep1}, \rf{tdep2}.  For  tachyons we get some functions $M_1^2(t)$ and for masses which are not tachyonic at small $t$ but become tachyonic away from small $t$ we get some functions $M_2^2(t)$. We plot both functions for $0.001 \lesssim t \lesssim 1.3$ in Fig. \ref{Masses}. 
By looking at the plots in Fig. \ref{Masses} we can see that the numerical value of the tachyonic mass squared $ M_1^2(t) $ in the limit $t\to 0$ becomes infinitely large and negative. At large $t$ it becomes exponentially large and negative.  Moreover, the absolute value of the tachyon mass square is always very large,  even at its minimum value at $t\approx 0.27$ where it is given by
 $|M_1^2(t)| \gtrsim 3 \cdot 10^3$.

For the second eigenvalue $M_2^2(t)$ at small $t$ the mass squared is positive and grow infinitely large at $t\to 0$, but at  $t \sim 0.34$ it changes its sign, and becomes exponentially large and negative. 

The dimensionless values of the masses in eq. \rf{tach1} take these numbers above. The question is: What are their dimensionful values? In \cite{DallAgata:2022abm} it is suggested that the dimensionful couplings come in units of $\mu$, therefore we may assume that the same holds for the masses. In fact, there are only 2 mass squared parameters in this construction: $\mu^2$ and  $\Lambda^2$ as the authors of \cite{DallAgata:2022abm} use $\xi_{UV}= {f\over \Lambda^2}=1$.

At small $t \to 0$ we have $\mu \to \Lambda$, at $t =2$ we have $\mu \to e^{-2}\Lambda$ and the values of the masses squared are
\bea
&\quad \! -M_1^2 \Big |_{t=0.001} \sim 3\times 10^{5} \Lambda ^2\ , \qquad &-M_1^2 \Big |_{t=2} \sim 5 \times 10^{3} \Lambda ^2\ ,\cr
& M_2^2 \Big |_{t=0.001} \sim 500 \Lambda ^2\ , \qquad &-M_2^2 \Big |_{t=2} \sim 50 \, \Lambda ^2\ .
\eea
At the minimum at $t=0.27$ $\mu \to e^{-0.27}\Lambda$ we find
\be
-M_1^2 \Big |_{t=0.27} \sim 1750 \, \Lambda ^2 \ .
\ee

There is a discussion in \cite{DallAgata:2022abm} that the ERG evolution cannot really be trusted for $t \gtrsim 1$. However, we see now that the smallest $|M_1^2|$ at $t=0.27$ is a thousand times greater than the cut-off  $\Lambda ^2$ that was assumed to exist in the theory. At every other values of $t$ we see in Fig. \ref{Masses} that $M_1^2$ exceeds $\Lambda ^2$ even more. And in the small $t$ limit which is most interesting for us the masses become infinitely many times greater than $\Lambda $.

Thus, we are dealing with masses which are numerically either significantly greater or even many orders of magnitude greater than the cut-off scale $\Lambda^2$. This raises questions about the consistency of the set of assumptions on which the setup of the ERG equations is based, and suggests that this approach is completely inapplicable in the limit $t\to 0$, where its results are supposed to match the results of the VA model.

In supergravity it is suggested in \cite{DallAgata:2022abm} that $\Lambda ^2 \sim 10^{-8} M_{Pl}^2$ for $e^{2t} \sim {\cal O} (1)$. i.e. 8 order of magnitude below the Planck mass squared. It means that the smallest value of $|M_1^2|$ is $10^3 \Lambda^2 \sim 10^{-5} M_{Pl}^2$. Clearly, the cut-off assumption is inconsistent with these numbers.

\section{Instabilities in string theory?}\label{sec:string}
In this section we would like to revisit the claim of an anti-D3-brane instability \cite{DallAgata:2022abm} from the point of view of the string theory KKLT setup. To that end let us quickly review the related developments in recent years.

Given that the background fluxes in the KKLT construction carry charge with the opposite sign as the anti-D3-brane(s), the backreaction leads to an accumulation of flux near the anti-D3-brane. One might then worry that this might trigger an immediate annihilation of the anti-D3-brane against the fluxes (via the decay channels discussed in the probe limit by Kachru, Pearson and Verlinde (KPV) \cite{Kachru:2002gs}). Furthermore, the flux accumulation seemed to lead to nonphysical singularities in the flux background. However, it was recently discovered through careful analysis that neither of these problems really manifest \cite{Michel:2014lva, Gautason:2015tla, Cohen-Maldonado:2015ssa, Blaback:2019ucp}. In particular, \cite{Blaback:2019ucp} reviews how anti-brane polarization resolves flux singularities and leads to valid supergravity solutions for sufficiently large numbers of anti-branes.

The case of a small number of anti-branes or actually a single anti-brane (potentially placed on top of an O-plane) has also been studied extensively in the last few years \cite{Michel:2014lva, Kallosh:2014wsa, Bergshoeff:2015jxa, Kallosh:2015nia, Bandos:2015xnf, Garcia-Etxebarria:2015lif, Dasgupta:2016prs, Vercnocke:2016fbt, Kallosh:2016aep, Bandos:2016xyu, Aalsma:2017ulu, GarciadelMoral:2017vnz, Aalsma:2018pll, Cribiori:2019hod}. This is particularly important for our discussion here since in this case the connection between the VA action for the goldstino and the anti-D3-brane worldvolume fields has been made manifest (see also \cite{Parameswaran:2020ukp} for a generalization to multiple anti-branes). So, we will from now on mostly focus on the case of a single anti-D3-brane.\footnote{It was shown in \cite{Bena:2014jaa} that multiple anti-D3-branes seem to repel each other. Given that they are confined to the bottom of a warped throat in KKLT type constructions, this might then justify the study of each of them as a single anti-D3-brane, even when multiple anti-D3-branes are present \cite{Michel:2014lva}.} For a single anti-brane in a flux background the authors of \cite{Michel:2014lva} argue that one should include the anti-D3-brane action (that contains the VA action) as part of the low energy field theory, which is what has been done so far in the literature and also in \cite{DallAgata:2022abm}. However, the authors of \cite{Michel:2014lva} find no instability in this setup and show that divergences in the flux cloud near the anti-brane get resolved by matching onto string theory at short distances. Furthermore, the authors of \cite{Michel:2014lva} show that the \emph{only} allowed anti-D3-brane instability is the NS5-brane instanton described by KPV \cite{Kachru:2002gs}. As mentioned above this decay channel has been studied in great depth and in particular from the stand point of non-linear supersymmetry in \cite{Aalsma:2017ulu, Aalsma:2018pll}. There exists now universal agreement in the literature that there are metastable vacua after the polarization of the anti-D3-branes into an NS5-brane. As stated in \cite{DallAgata:2022abm}, the connection of their instability to the KPV paper is unclear, so it could be that their result is related to the polarization of anti-D3-branes into a metastable NS5 brane. However, at least for a single anti-D3-brane the claims in \cite{DallAgata:2022abm} seem to be at odds with the perturbative stability that was observed in \cite{Michel:2014lva} via a more explicit study of an anti-D3-brane in a flux background.

The explicit connection between an anti-D3-brane and the VA action was first made in~\cite{Kallosh:2014wsa}, where it was shown that the low energy effective action for an anti-D3-brane on top of an O3-plane is actually given by four copies of the VA action. This setup is particularly simple (since it is in flat space without fluxes) and it was shown to be stable by an explicit string theory calculation in \cite{Uranga:1999ib}. In \cite{Bergshoeff:2015jxa} this setup was generalized by including background fluxes and it was found that these give masses to three of the four copies of the VA fermions, leaving at low energies essentially only the VA action. It was then shown that such a setting of an anti-D3-brane on top of an O3-plane can arise in the warped throats that are required for the KKLT construction \cite{Kallosh:2015nia, Garcia-Etxebarria:2015lif}. The full action of an anti-D3-brane in a warped throat (that is \emph{not} placed on top of an O3-plane) was worked out in \cite{GarciadelMoral:2017vnz, Cribiori:2019hod}. Each of these previous described works that study different variants of the anti-D3-brane action in backgrounds, varying from flat space to KKLT flux compactifications, contain the goldstino and the VA action as well as other light fields. 

The authors of \cite{DallAgata:2022abm} start from a particular realization of the VA action imposed via a Langrange multiplier field $T$ that then becomes dynamical and ultimately leads to an instability. If that argument were to carry over to the anti-D3-brane and in particular to the anti-D3-brane uplift in the KKLT scenario, then one has to wonder whether it equally carries over to other scenarios as well: Four copies of the VA action arise from the worldvolume fermions on an anti-D3-brane on top of an O3-plane \cite{Kallosh:2014wsa}. If we remove the O3-plane, the fermionic action is unaltered \cite{Kallosh:1997aw} and supplemented with a U(1) gauge field and three complex scalars. In flat space, however, there is no distinction between an anti-D3-brane and a D3-brane. This means these worldvolume fields on the brane form an $\mathcal{N}=4$ vector multiplet and the action is invariant under 16 linearly and 16 non-linearly realized supersymmetries. Clearly this supersymmetric system of an $\mathcal{N}=4$ vector multiplet is stable, although the action for the fermions is of VA type \cite{Kallosh:1997aw}. Now let us do an O3 orientifold projection again but place the anti-D3-brane away from the orientifold (that maps it to a mirror anti-D3-brane). The action for this anti-D3-brane in flat space is exactly the same as the $\mathcal{N}=4$ vector multiplet action on the D3-brane. The orientifold background only projects out the 16 linear supersymmetries but it does not project out any of the fields on the anti-D3-brane. So, as is common in (intersecting) brane models in string theory, locally the amount of preserved supersymmetry is larger ($\mathcal{N}=4$ in 4d) than globally ($\mathcal{N}=0$ in this case). This simple thought experiment shows that it is nonsensical to conclude from the sheer presence of the VA action that a vacuum is unstable. The completion of the VA action through other fields leads for D3-branes and anti-D3-branes to locally supersymmetric actions. Any potential instability can only arise via the coupling to the background fields that requires a careful study and has been performed for the anti-D3-brane in the KKLT setup in the existing string theory literature, reaching the conclusion that the system is (meta-)stable.

The above string theoretical argument carries over to for example any 4d globally supersymmetric theory: If in a vacuum four supercharges are spontaneously broken, then there is a massless goldstino whose action is given by the VA action.\footnote{If all other fields are massive, then this is actually the entire action of the theory at sufficiently low energies.} If the pure VA model has an instability, then this would either render any \emph{partially} SUSY breaking vacuum in \emph{any} SUSY theory unstable, which seems absurd, or it does not necessarily imply a problem once other fields are included and therefore requires a model dependent study. Such a model dependent study for the KKLT model was not performed in \cite{DallAgata:2022abm} for the anti-D3-brane uplift, so no conclusion regarding its stability can be reached based on the analysis in \cite{DallAgata:2022abm}. In particular, for the anti-D3-brane in the KKLT scenario the SUSY breaking scale is set by the string scale at the bottom of the warped throat. This string scale is above the (warped down) KK scale so in addition to the light open and closed string states with masses below the warped down string scale, there are also many states from the KK tower that are lighter than the SUSY breaking scale. Furthermore, for a single anti-D3-brane at the bottom of a warped throat there is a massless U(1) gauge field \cite{GarciadelMoral:2017vnz, Cribiori:2019hod} in addition to the massless goldstino. It is therefore impossible to remove all light fields from the theory by reducing the UV cutoff. Therefore, one cannot apply the results of \cite{DallAgata:2022abm} to the anti-D3-brane uplift in KKLT. 

The concern of the previous paragraph is briefly addressed at the end of the introduction in~\cite{DallAgata:2022abm} were the authors say: ``The only way to avoid the instability \ldots~would be to always have some additional light states". If one identifies the UV scale with the supersymmetry breaking scale, then there seem to be \emph{generically} additional light states at or below this scale. If one deals with a vacuum in which \emph{all} fields but the goldstino are massive, then one could potentially choose the UV scale to be below the mass of the lightest states and the theory reduces to only the VA model. However, this would at best apply to an anti-D3-brane on top of O3-plane and not to the generic anti-D3-brane uplift of KKLT since there is a massless U(1) gauge field in addition to the goldstino.


\

The supergravity model used in \cite{DallAgata:2022abm} to discuss the KKLT model is the model in their Sec. 4.2 with two propagating unconstrained superfields $X, T$ in addition to a superfield $\rho $ representing the volume of the extra dimensions. The KKLT supergravity model in $M_{Pl}=1$ units is \cite{Ferrara:2014kva}
\be
K = -  3\ln (\rho+\bar \rho) +X^2\, ,  \qquad W= fX + W_0 + A e^{ -a\rho}\, , \qquad X^2 =0 \ .
\ee
This model is replaced in \cite{DallAgata:2022abm} by the one where there is also a superfield $T$ which is a propagating superfield. Namely, the total \K and superpotential include the terms in eqs. \rf{Ka} and \rf{Su} as well as $\Delta K = -  3\ln (\rho+\bar \rho)$ and $\Delta W= W_0 + A e^{ -a\rho}$. They include some $t$-dependent factors of the form $e^t$ which at small $t$ are  regular.
Most importantly, the superfield $X$ is not nilpotent anymore, since $T$ is a dynamical superfield.

The presence of $W_0$ leads to a shift of the saddle point of the potential to some non-vanishing values for the scalar fields. We have computed the masses and the potential in this more general model, only to confirm that at small $t$ eqs. \rf{tach1lim1} and \rf{small_t} are beyond repair: all the additional terms in $K$ and $W$ are regular at small $t$ and cannot remove the singularities in these equations. 

It is also interesting that Fig. 5 in \cite{DallAgata:2022abm} shows the numerical example of the KKLT volume field potential at $t=0.1, 0.3, 1$ where the authors of \cite{DallAgata:2022abm} find that at $t= 0.3, 1$ the volume field potential is negative at the critical point but at smaller $t=0.1$ the minimum is still in de Sitter. But the $t\to 0$ limit remains not well defined and singular in the $X,T$ sector, which means that the anti-de Sitter minimum for the volume at $t= 0.3, 1$ is not relevant to the model with a nilpotent multiplet at $t=0$.

This explains a technical error in the statement in \cite{DallAgata:2022abm} about the tachyonic instability in the KKLT model. The tachyonic instability takes place in the model studied in \cite{DallAgata:2022abm}, however, the limit from this model to the KKLT model involves a singularity and therefore these are two different models.

\section{Discussion}

The proposal in \cite{DallAgata:2022abm} is to modify the Volkov-Akulov theory \cite{Volkov:1973ix} in the form given in \cite{Komargodski:2009rz} which we present here in eq. \rf{VA1}. In \cite{DallAgata:2022abm} they have introduced a kinetic term for the Lagrangian multiplier   superfield $T$ by adding a term to the \K potential such that $T$ instead of being a Lagrange multiplier becomes a propagating field
\be
K \rightarrow K +Z_T \, T\bar T    \ .
\ee
The analysis in  \cite{DallAgata:2022abm}  after they make the Lagrange multiplier a propagating field, proceeds along the lines of the non-perturbative ERG equations. They were following \cite{Jaeckel:2002rm}  where the bound states in a model with a single fermion were studied  and a kinetic term for the  auxiliary scalar was added to the fermion action {\it ad hoc}. In  \cite{Jaeckel:2002rm} and \cite{DallAgata:2022abm} the ${1\over N}$ approximation used normally for studies of fermion condensates is not available.

The narrative in \cite{Jaeckel:2002rm} where a kinetic term $Z_\phi \partial_\mu \bar \phi \partial^\mu  \phi $ was added to the model, required the study of the limit $Z_\phi \rightarrow 0 $ of the solutions of the ERG equations. The authors of  \cite{Jaeckel:2002rm, Braun:2011pp}  argued that their model with  bosonization is equivalent to the original fermion model. They explained it as follows:  
`the corresponding bosonic species becomes very massive and therefore effectively drop out of the flow'.

In \cite{DallAgata:2022abm} the renormalization group equations were solved at   $Z_T \not = 0$. The model has 2 complex scalar fields, the first components of the unconstrained superfields $X$ and $T$. A change of variables 
\be
Z_T\, T \bar T\rightarrow   T { \bar T} \ , \qquad {1\over 2}TX^2  \rightarrow   {1\over 2 \sqrt {Z_T} }TX^2 \ ,
\ee
converted the former Lagrange multiplier into a canonical propagating field.  It also made the coupling between $T$ and $X$ proportional to $ {1\over  \sqrt {Z_T} }$. This feature of the coupling is the reason for the problem: The model in \cite{DallAgata:2022abm} encounters a singularity on the way  back to $Z_T=0$. If this singularity would be only in the sector of scalar fields with positive mass squared, it might work well towards the limit to the original VA  fermionic theory as these heavy scalars would decouple from the spectrum. But such a singularity  in the tachyonic sector of the theory shows that the theory is not related to a fermionic model since there is an instant vacuum instability with respect to the generation of classical scalar fields, instead of decoupling.

The detailed form of the model in \cite{DallAgata:2022abm} is given in our eqs. \rf{Ka}, \rf{Su}. The model has two unconstrained superfields with canonical kinetic terms at $X=T=0$. The couplings $\tilde \gamma, \tilde \zeta, \tilde f, \tilde g$ depend on the renormalization group time $t= \ln {\Lambda\over \mu}$. There is a claim in  \cite{DallAgata:2022abm} that in the limit $t=0$ their model in the global supersymmetric case is the Volkov-Akulov model, and in the local supersymmetric case it represents supergravity interacting with the nilpotent multiplet. However, this claim is not justified for several reasons: 

\noindent 1) In the limit $t\to 0$ ($Z_T \approx {1\over 16 \pi^2}  t^2 \to 0$) two of the couplings between the canonical scalars in  eqs. \rf{Ka}, \rf{Su} blow up,
\be
\tilde \gamma  \rightarrow {32 \pi^2\over   t}\, , \qquad 
\tilde g \rightarrow {2\pi\over   t} \ .
\label{small}\ee  
2)   The masses of  scalars  at $X=T=0$ were computed and tachyons were discovered in \cite{DallAgata:2022abm}. 
However we have found that the limit of negative mass squared is singular at
 $t\to 0$. This leads to an instant vacuum decay and shows that the corresponding scalar fields are not decoupled in this limit.
 
 \noindent 3) We have computed
the scalar potential at small $t$ in the case where all scalars do not vanish. We have found   that this potential has terms diverging  polynomially in the global case and exponentially in supergravity
\be
V |_{t \to  0} =  {F_1 \over  t^3}  e^{F_2\over t}   \ .
\label{smalt}\ee
Here $F_1, F_2$ are functions of the scalar fields which are not vanishing when the scalar vevs are not vanishing, see eq. \rf{small_t}.

 \noindent 4) The part of the action quadratic in fermions vanishes at vanishing values of the scalars,  but it blows up at small $t$ when the scalar fields have non-zero values. The matrix $m_{ij}$ in the action \rf{ferm}  has one vanishing eigenvalue and one non-vanishing eigenvalue singular at $t\to 0$,
\be
m_f |_{t \to  0}\sim {F_3\over t^3}    e^{  F_4 \over  t} \ .
\label{fermC}\ee
Here $F_3, F_4$ are functions of the scalars which are not vanishing when the scalars are non-vanishing, see eq. \rf{eig}.

 \noindent 5) In the case of $X=0$ we studied the potential of the $T$ field and we have found a complex flat direction at small $t$, shown in Fig. \ref{Tpotential}. This corresponds to  a massless complex scalar  field, which is absent in the VA theory.

In addition to conceptual problems concerning a  relation between  the 2-superfield  model in \cite{DallAgata:2022abm} to the VA theory in the UV limit $\mu \to \Lambda$ there is a significant internal problem within the 2-superfield field model in \cite{DallAgata:2022abm}. Namely, the scale $\Lambda^2$ is declared to be a UV cut-off so that only scales below this cut-off scale are relevant. However, as we have shown in Sec. \ref{sec:num} the masses squared of the tachyons and normal scalars at $t \to 0$ become infinitely large. At small $t\sim 10^{-3}$ with  $\sqrt f  \sim \Lambda = \mu (1+10^{-3})$ 
\be
|M^2_{tachyons} | \sim 10^5 \Lambda^2\, ,  \qquad |M^2_{normal} | \sim 10^2 \Lambda^2 \ .
\label{masses}\ee
The smallest value of $|M^2_{tachyons} |$  is at $t\sim 0.27$, and at larger $t$ it grows exponentially, so  
\be
|M^2_{tachyons} | \gtrsim 10^3 \Lambda^2 \ ,
\label{min}\ee
for the full range of $t$ along the ERG trajectory. These masses are even greater in the regime $f > \Lambda^2$ considered in \cite{DallAgata:2022abm}.
This makes the whole setup of the ERG in the theory in \cite{DallAgata:2022abm} questionable, regardless of its relation  to theories with non-linear supersymmetry.   In this evaluation of masses we are using their own eq. (3.7)  in \cite{DallAgata:2022abm}. But the Polchinski flow equation \cite{Polchinski:1983gv,Litim:2018pxe} 
 for a Wilsonian action approach  is based on a restriction on the allowed momenta of the form
\be
p^2 \leq \mu^2 \leq \Lambda ^2 \ .
\ee
This is not consistent with masses they have found,  as we show in eq. \rf{masses}, \rf{min}. Thus the statement in  \cite{DallAgata:2022abm} that they have found a self-consistent ERG flow clashes with the values of the scalar masses they have found but did not compare with their cut-off scale $\Lambda^2$. 
 
We conclude that the supersymmetry/supergravity model in eqs. \rf{Ka}, \rf{Su} with two unconstrained superfields  studied in \cite{DallAgata:2022abm} is, indeed, unstable at vanishing scalars. However, the  limit when the theory studied in  \cite{DallAgata:2022abm} approaches the Volkov-Akulov theory is discontinuous. For example, at $X=0$ there are 2 massless scalars, which are absent in the VA theory. Therefore, we do not see any evidence that  the  model developed in \cite{DallAgata:2022abm}   represents the VA global non-linear supersymmetry model \rf{VA}. We found that the same conclusion is valid for the 
de Sitter supergravity model \rf{finalactionf1} and the KKLT model. Moreover, it is doubtful that the ERG flow approach \cite{Polchinski:1983gv,Litim:2018pxe}  proposed in  \cite{DallAgata:2022abm} is consistent in view of the fact that in both in the UV limit and in the IR limit it describes  states with masses significantly above the UV cut-off. We conclude that
 the assertion in \cite{DallAgata:2022abm} of an instability of de Sitter vacua in  theories with non-linear realization of supersymmetry is not substantiated by their investigation.

Moreover, concerning the arguments of an instability of the KKLT string theory construction, we have explained in Sec. \ref{sec:string} that one cannot reach such a conclusion based on the analysis performed in \cite{DallAgata:2022abm}. This argument is completely independent of the discussion above and relies on the simple observation that the VA theory arise at low energies for \emph{any} theory with a (partially) supersymmetry breaking vacuum. Clearly not all such theories have unstable vacua. The loophole in the analysis presented in \cite{DallAgata:2022abm} is stated at the end of their introduction. The authors assume the absence of light states below their cut-off scale. Given that the SUSY breaking scale in the KKLT scenario is set by the warped down string scale, which is above the warped down KK scale, there are many additional light fields in the KKLT construction. In particular, a single anti-D3-brane has among its worldvolume fields a massless U(1) gauge field \cite{Cribiori:2019hod}. Such light modes were not taken into account in \cite{DallAgata:2022abm} and therefore one cannot claim that the work in \cite{DallAgata:2022abm} applies to the anti-D3-brane in the KKLT scenario.

\section*{Acknowledgement}
We are grateful to  E. Bergshoeff, S. Ferrara, D. Freedman,  A. Karlsson, D. Murli and A. Van Proeyen, our collaborators  on the earlier work where the theories with non-linear supersymmetry were investigated. RK is grateful to D. Volkov for explaining his work a long time ago in Kharkiv, Ukraine. The work of RK and AL is  supported by the SITP and by the US National Science Foundation Grant PHY-2014215. The work of TW is supported in part by the NSF grant PHY-2013988. The work of YY is supported by a Waseda University Grant for Special Research Projects (Project number: 2022C-573).

\appendix

\section{From linear to non-linear supersymmetry in a simple example}\label{appA}

The VA theory \cite{Volkov:1973ix} can be presented in the form of a nilpotent chial superfield $X^2=0$ \cite{Komargodski:2009rz} 
\be
 {\cal S}_{VA-KS} = \int d^4 x \ls \int d^4\theta K + \lp\int d^2 \theta (W  + c.c\rp \rs \ ,
\label{VAKS}\ee
where
\be
K= X\bar X\, ,\qquad W= f X\, ,\qquad X^2=0 \ .
\label{KW}\ee
Solving the constraint equations $X^2=0$ one finds that the superfield depends only on the goldstino $G$ and on a constant $F$, so that $X={G^2\over 2F} + \sqrt 2 \,\theta G + \theta^2 F $ and one can identify the action with a non-linearly realized supersymmetry in the form given in  \cite{Komargodski:2009rz} \be
{\cal L}_{VA-KS}=-f^2 +i \partial_\mu  \bar G \bar \sigma^\mu G + {1\over 4 f^2} \bar G^2 \partial^2 G^2 - {1\over 16 f^6} G^2  \bar G^2 \partial^2 G^2 \partial^2 \bar G^2 \ .
\label{VA}\ee

The supergravity version of this theory was constructed in \cite{Bergshoeff:2015tra,Hasegawa:2015bza}. We display it for the convenience of the reader in Appendix~\ref{appC}. We present there an action which is invariant under a non-linear local supersymmetry and that is called de Sitter supergravity. In the global case, the action depends  on only the goldstino and is given in eq. \rf{VA} above. In the local case, the action depends on the goldstino, the graviton and the gravitino and it is given in eqs. \rf{finalactionf1}-\rf{Dmupsinu} below. The non-linear supersymmetry transformations are presented in eqs. \rf{delte}- \rf{deltgold}. Moreover, in the de Sitter supergravity due to local supersymmetry one can take a gauge where the goldstino is vanishing.

We explain now why some linear supersymmetry models are related to non-linear supersymmetry ones, and some are not. In an example studied in  \cite{Komargodski:2009rz} it was shown how to get from linear SUSY to constrained superfields. The proposal in \cite{Komargodski:2009rz} is to start with the theory with an unconstrained single superfield $\Phi$ with linear supersymmetry and a canonical \K potential $K= \Phi\bar \Phi$ with $W= f \Phi$. The scalar field here is massless. One can add corrections to the \K potential which can be motivated, e.g. by the 1-loop computation in the O'Raifeartaigh model \cite{ORaifeartaigh:1975nky} (see details in the next subsection)
\be
K= \Phi\bar \Phi - {c\over M^2} \Phi\bar \Phi \Phi \bar \Phi\, , \qquad W= f \Phi\, , \qquad  c>0 \ ,
\label{KS} \ee
\be
\Phi = \phi+ \sqrt 2 \theta \psi_{\phi} + \theta^2 F_{\phi} \ .
\ee
Such a theory can arise as the low-energy Lagrangian below some scale $M$ after neglecting higher order terms in ${1\over M}$. It is valid for energies $\sqrt f\ll E\ll M$. The scalar potential of this theory is
\be
V= {f^2\over 1-{4c\over M^2} \phi\bar \phi} \approx f^2 \Big [1+ {4 c\over M^2} \phi\bar \phi +  \dots \Big ] \ ,
\label{V}\ee 
where terms small at large $M$ are neglected. 
There is a   minimum at $\phi=0$ and the mass of the complex scalar $\phi$ is
\be
m^2_\phi = 4c {f^2\over M^2} >0  \ .
\label{mass}\ee
At this point things depend crucially on the sign of the constant $c$. The minimum of this potential near $\phi=0$ requires that $c$ is positive. This was a suggestion in  \cite{Komargodski:2009rz}, and in such case one finds that $m^2_\phi = 4c {f^2\over M^2} > 0$, the minimum at $\phi=0$ is stable. If the sign is negative, the point  $\phi=0$ is unstable. 

We can see exactly the same example in eq. (3.2) in \cite{Cribiori:2017ngp}  where $c$ in eq. \rf{KS}  is taken to be  $+1$ and $M^2$ is called $\Lambda^2$.  However, in the recent paper \cite{DallAgata:2022abm} two of the authors of  \cite{Cribiori:2017ngp} argued that the sign of $c$ in eq. \rf{KS} has to be negative for the consistency of their ERG equations. 

To explain the importance of this step, we recall how the reasoning of  \cite{Komargodski:2009rz,Kallosh:2015pho} works in the context of eq. \rf{KS} with $c>0$ and why it explains how to go from linear SUSY to constrained superfields. For example, in \cite{Komargodski:2009rz} they have integrated out massive bosons with the mass given in eq. \rf{mass}, and at small momenta in the IR they found that 
\be
{\cal L} = - f^2 +|F_\phi + f|^2 - {c\over M^2} |2\phi F_\phi- \psi^2|^2 +\dots
\ee
Here $\dots$ stand for terms with derivatives. They have solved the equations of motion in the IR where neglected terms are small in their approximation and concluded that 
\be
\phi = {\psi^2\over 2F_\phi}\ ,  \qquad \Phi^2=0 \ , \qquad  \Phi \sim X\, ,    \qquad \Rightarrow \qquad X^2=0 \ .
\ee
This derivation of the constrained superfield in the IR starting from the unconstrained superfield\footnote{This procedure was generalized in \cite{Kallosh:2015pho} where the complete VA action was derived, including higher derivative terms, starting with eq. \rf{KS}.}
 would  be invalid for $c<0$, where the scalars near $\phi=0$ are tachyonic. 
This means that  the  model with linear supersymmetry in eq. \rf{KS} with negative $c$ is unstable at small $\phi$. But it also means that it is not related to the VA theory. This is opposite to the case with positive  $c$, which is related to the VA theory, as shown in \cite{Komargodski:2009rz,Kallosh:2015pho}.

The 1-loop computation in the O'Raifeartaigh model leads to an effective quartic term in the \K potential in \rf{KS} with $c>0$, see for example \cite{Komargodski:2009rz}. The original detailed computation was done in    \cite{Huq:1975ue} and the positive value of $c$ was confirmed in  \cite{Witten:1981kv,Intriligator:2007py}.
These results for the O'Raifeartaigh model were applied in  the context of O'KKLT supergravity in 
\cite{Kallosh:2006dv}.

The classic O'Raifeartaigh superpotential \cite{ORaifeartaigh:1975nky} involves three superfields $\phi_1, \phi_2, X$ with the superpotential in the notation of \cite{Intriligator:2007py}
\begin{eqnarray}
W = m \phi_1 \phi_2 + {h\over 2} X \, \phi_1^2 + f X 
\end{eqnarray}
and canonical K\"{a}hler terms.  There is also a $\mathbb{Z}_2$ symmetry under which $\phi_1$ and  and $\phi_2$ are odd. 

The 1-loop  potential  was computed in \cite{Huq:1975ue} as a function of $X$. The value of $X$ is not fixed at tree level, but the 1-loop contribution lifts this flat direction through contributions from bosons and fermions in the Coleman-Weinberg \cite{Coleman:1973jx} type potential depending on $X$
\be
V_{\rm eff} = V_{\rm tree} + {1\over 64 \pi^2} {\rm STr}  {\cal M}^4( X) \ln {{\cal M}^2( X)\over M^2_{\rm cutoff}} \ .
\label{CW}\ee
The total 1-loop potential in the O'Raifeartaigh model is complicated but explicitly given in  \cite{Huq:1975ue}, see eqs. (4.7)-(4.9) there. This potential  was shown to have a minimum at $X=0$.

In \cite{Intriligator:2007py} it was also found that for $y=\Big |{hf\over m^2} \Big |<1$ the vacuum of the classical theory has the $\mathbb{Z}_2$ symmetry unbroken. For $y>1$, the $\mathbb{Z}_2$ symmetry is broken, and $X$ is arbitrary in both cases. 
 This degeneracy is lifted by the 1-loop Coleman-Weinberg potential so that the potential at small $X$ is
 \be
V_{\rm eff}(X) = V_0 + m_X^2 |X|^2 + {\cal O} (|X|^4) \ ,
\label{eff} \ee
\be
m_X^2= {1\over 32 \pi^2} \left|h^2 m^2\right| \hat f_a(y) >0 \, , 
\label{massS}\ee
where the function $ \hat f_{a=1}(y)$ is positive for $y<1$:
\be
 \hat f_1(y)\equiv y^{-1} \Big((1+y)^2 \ln(1+y) - (1-y)^2 \ln (1-y) -2y\Big) \, , \qquad    y= \left|{hf\over m^2} \right|<1 \ .
\ee
Thus, the positive mass formula derived by Huq \cite{Huq:1975ue} back in 1975 was confirmed. The second phase with the $\mathbb{Z}_2$ symmetry broken was discovered in \cite{Intriligator:2007py} and also has a positive mass squared. The function $ \hat f_{a=2}(y)$ in eq. \rf{massS} is positive for $y>1$
\be
 \hat f_2(y)\equiv y^{2} \ln y - (y-1)^2 \ln (y-1)- (y-1/2) (2\ln (y-1/2) +1)  \, , \qquad y= \left|\frac{hf}{m^2} \right|>1 \ .
 \ee
Higher loop corrections are suppressed by powers of $h^2$.

One can interpret this 1-loop corrected potential $V_{\rm eff}$ in eq. \rf{eff} at small $X$ as the expression in eq. \rf{V} coming from the corrected \K potential \rf{KS}. There is no doubt in this case that eq. \rf{KS} with positive $c$ is an example of the 1-loop quantum correction of the O'Raifeartaigh model. This was also a choice in  eq. (3.2) in \cite{Cribiori:2017ngp} since at that time the authors were interested in the transition `From Linear to Non-linear SUSY'.

Thus, the \K potential stabilizes $X$ at the origin. If one decides to apply an RG approach here and resum the logs one can see that this leads to a Landau pole and creates problems for the physics of this model at large $X$. To avoid the Landau pole and make the ERG approach consistent, one has to make $m_X^2$ tachyonic. This was the choice in \cite{DallAgata:2022abm} where, to avoid the Landau pole at all $X$ and based on the ERG equations, the authors studied a theory that is unstable at small $X$.

Moreover, it was  observed in \cite{Witten:1981kv, Intriligator:2007py} that in more general supersymmetric models where also non-abelian gauge couplings are added at large $|X|$ the 1-loop potential behaves as 
\be
V_{\rm eff}(X) \to  (c_h h^2- c_v g^2) \ln \ls {|X|^2\over M^2_{\rm cutoff}} \rs\, , \qquad c_h>0,\quad c_g >0 \ ,
\ee
where $h$ is a Yukawa coupling and $g$ is a non-Abelian gauge coupling. In the absence of non-Abelian couplings the theory with scalars and fermions is known to have a positive term $c_h h^2$ in front of the $\ln |X|^2$. An opposite choice is now taken in eq. (3.1) in \cite{DallAgata:2022abm} where the constant $\tilde \zeta = -{c\over M^2}$ is positive, i.e. $c$ in eq. \rf{KS} here is negative. 

To conclude, the 1-loop quantum corrections of the O'Raifeartaigh model show that this case is an example of a model with linear supersymmetry which can be related to a model with a non-linear supersymmetry. In the opposite case of a linear supersymmetry model with $c<0$ at small $X$ the model is unstable, but also not related to the VA theory. The bridge between the linear and non-linear theories involves integrating out the scalars, which is only possible for $c>0$.

\section{Pure de Sitter supergravity }\label{appB}
The action invariant under spontaneously broken local supersymmetry depends on the vierbein and gravitino from the gravitational multiplet $e^a_\mu,~\psi_\mu$ and on the goldstino $\chi$ from the nilpotent multiplet \cite{Bergshoeff:2015tra,Hasegawa:2015bza}: 
\begin{eqnarray}
 e^{-1}{\cal L} & = &\frac{1}{2\kappa^2} \left[ R(\omega (e )) -\bar \psi _\mu \gamma ^{\mu \nu \rho } D^{(0)}_\nu \psi _\rho
 +{\cal L}_{\rm SG,torsion} \right] +3 \frac{m^2}{\kappa^2} - f^2 \nonumber\\
&&+\frac{f}{\sqrt{2}} \bar \psi_\mu \gamma
  ^\mu   \chi+\frac{m}{2\kappa^2}  \bar \psi _{\mu } \gamma ^{\mu \nu }\psi _{\nu } +\frac{\kappa^2}{24} \chi ^2\bar \chi ^2\nonumber\\
 &&-\ft12\bar \chi  \slashed{D}^{(0)}\chi -\ft{1}{32}\rmi\,e^{-1}\varepsilon ^{\mu \nu \rho \sigma }\bar \psi _\mu \gamma _\nu \psi _\rho \bar \chi \gamma _*\gamma _\sigma \chi-\ft12\bar \psi _\mu P_R\chi \bar \psi ^\mu P_L\chi\nonumber\\
&& +\frac{\bar  \chi^2}{2 f}A \frac{\chi^2}{2 f} -\left(\frac{ \chi^2}{2 f} \bar B+\frac{\bar \chi^2}{2 f} B\right) - \frac{\chi^2\bar \chi^2}{16 f^4}\left(\frac{\bbox\chi^2}{f}-2B\right)\left(\frac{\bbox\bar \chi^2}{f}-2\bar B\right)\,,\label{finalactionf1}
\end{eqnarray}
where
\begin{eqnarray}
\chi ^2&\equiv& \bar \chi P_L\chi \,,\qquad   D^{(0)}_\mu    = \partial _\mu +\ft14 \omega _\mu {}^{ab}(e )\gamma _{ab}\,,\nonumber\\
  {\cal L}_{\rm SG,torsion}&=&-\ft{1}{16}\left[
(\bar{\psi}^\rho\gamma^\mu\psi^\nu) ( \bar{\psi}_\rho\gamma_\mu\psi_\nu
+2 \bar{\psi}_\rho\gamma_\nu\psi_\mu) - 4 (\bar{\psi}_\mu
\gamma\cdot\psi)(\bar{\psi}^\mu \gamma\cdot\psi)\right]\nonumber\\
&&-e^{-1}\partial _\mu \left(e\bar \psi \cdot \gamma \psi ^\mu \right)\ ,
\label{covder0}
\end{eqnarray}
\begin{equation}
  A= \bbox + \rmi t^\mu \partial _\mu + \ft12\rmi e^{-1}\partial _\mu (e\,t^\mu) + r\,, \qquad
  \bbox =\frac{1}{\sqrt{g}}\partial _\mu \sqrt{g}g^{\mu\nu}\partial _\nu\,,
 \label{formA}
\end{equation}
\begin{eqnarray}
t^\mu &=&  \ft14 {\rmi} {\bar \psi}_\nu \gamma_* \gamma^{\nu\rho\mu}\psi_\rho\,,\quad
r =-\ft16 \left[R(\omega (e )) -\bar \psi _\mu \gamma ^{\mu \nu \rho } D^{(0)}_\nu \psi _\rho
    + {\cal L}_{\rm SG,torsion}
   -8\kappa ^2\,f^2\right] \,,\nonumber\\
 B  &=& \frac{1}{\sqrt{2}}\left[-e^{-1}\partial _\mu \left(e\bar \psi _\nu \gamma ^\mu \gamma ^\nu P_L  \chi  \right)-\ft23\bar \chi P_L \gamma ^{\mu \nu }D _\mu \psi _\nu  \right]
 + f
\Bigl(2 m +   
\frac{1}{2} {\bar\psi}_\mu\gamma^{\mu\nu}P_L\psi_\nu\Bigr)\,,\label{formB}
\end{eqnarray}
\begin{equation}
  D_\mu \psi _\nu=\left(\partial _\mu +\ft14 \omega _\mu {}^{ab}(e,\psi )\gamma _{ab}\right) \psi _\nu\,.
 \label{Dmupsinu}
\end{equation}
 All notations are explained in  \cite{Bergshoeff:2015tra}. 
The non-linear supersymmetry transformations of the fields $\chi$ and $e^a_\mu,~\psi_\mu$ are  the following: \\
For the fields of the gravity multiplet we
have
\begin{eqnarray}\label{delte}
\delta e^a_\mu &=& \frac12 \bar\epsilon \gamma^a\psi_\mu \ ,\\
\delta P_L\psi_\mu &=& P_L\bigg(\partial_\mu +\frac14 \omega_{\mu ab}(e,\psi)\gamma^{ab}-\frac32iA_\mu +\frac12i\gamma_\mu \slashed{A} +
\frac{\kappa}{2\sqrt3}\gamma_\mu \bar F^0 \bigg)\epsilon\label{deltpsi}\ ,
\end{eqnarray}
with
\begin{equation} 
  F^0=\overline{{\cal W}}_0= \sqrt{3}\,\frac{m}{\kappa}  +\frac{2}{\sqrt{3}} \kappa f\, X= \sqrt{3}\,\frac{m}{\kappa}- \frac{1}{\sqrt{3}}\kappa \chi ^2(1-{\cal A})\ ,
 \label{Fvalue}
\end{equation}
and
\begin{eqnarray}
A_\mu= \rmi\frac{\kappa^2}{6}\Big [ (\bar X\partial _\mu X -X\partial _\mu \bar X )  -\frac{1}{2}[\sqrt{2}\bar \psi _\mu(P_L\chi \bar X -P_R  \chi  X) +\bar \chi P_L\gamma_\mu \chi ]\Big ] \ ,
 \label{Amu}\end{eqnarray}
where
\be
X= - \frac{\chi^2}{ 2 f} (1-{\cal A}) \ , \qquad 
    {\cal A}= \frac{\bar \chi^2}{ 2 f^3} \left(A\, \frac{\chi^2}{2f} - B\right) \ .
 \label{calA}   \ee
The local supersymmetry transformation  for the goldstino is
\begin{equation}
 \delta P_L\chi =\frac1{\sqrt{2}} P_L\left[-f+(\slashed{\partial }-m)X
-  f {\cal A} \left(1-3\bar{{\cal A}}-\frac{\chi^2}{2f ^3}\bar B\right) \right]\epsilon
-\frac{1}{2}P_L\gamma ^\mu \epsilon \bar \psi _\mu P_L\chi
\,.
\label{deltgold}
\end{equation}

The action (\ref{finalactionf1}) is locally supersymmetric.   We now impose the unitary gauge condition that the goldstino is vanishing, $\chi=0$, and the action simplifies dramatically
\begin{equation}
 e^{-1}   {\cal L}_{\chi=0}
=\frac{1}{2\kappa ^2} \left[ R(e,\omega(e )) -\bar \psi _\mu \gamma ^{\mu \nu \rho } D^{(0)}_\nu \psi _\rho
 +{\cal L}_{\rm SG,torsion} \right] +\frac{3m^2}{\kappa^2} -f^2 +\frac{m}{2\kappa^2}\bar \psi _{\mu } \gamma ^{\mu \nu }\psi _{\nu }\,.
 \label{finalactionfFixed}
\end{equation}
For ${\bf  \Lambda} = f^2- 3m^2/\kappa^2 >0$ we have  a pure dS supergravity with a positive cosmological constant.
If some additional chiral unconstrained multiplets are present, a more general action of the type given in eq. \rf{finalactionf1} are also known, see for example \cite{Hasegawa:2015bza, Kallosh:2015tea, Schillo:2015ssx} in case of one or more chiral matter multiplets, in addition to a nilpotent one.

\section{Steps from Volkov-Akulov to the model proposed in \cite{DallAgata:2022abm}}\label{appC}

The KS model of a rigid supersymmetry \cite{Komargodski:2009rz} with 2 chiral superfields, $X$ and $T$, is
\be
 {\cal S} (X, T) = \int d^4 x \ls \int d^4\theta X\bar X + \lp\int d^2 \theta (fX + {1\over 2} TX^2) + c.c\rp \rs \ .
\label{SVA-KSxt}\ee
Here $X$ is satisfying the equation $X^2=0$, a nilpotency condition, as an equation of motion for the Lagrange multiplier superfield $T$. 

The proposal in \cite{DallAgata:2022abm} is to start with eq. \rf{SVA-KSxt} and add the following features to it: a cut-off $\Lambda$ and a suggestion that all couplings which are space-time constants depend on the `renormalization group time'  $
t_{RG}\equiv t= \log {\Lambda\over \mu} \geq 0
$
as a result of a solution of the ERG equations. Here $\mu$ is the  renormalization scale, related to the energy scale at which the theory is probed.

It is clearly explained in \cite{Komargodski:2009rz} that in the minimal case \rf{SVA-KSxt} the on-shell description of the theory with two superfields with linearly realized supersymmetry coincides with \cite{Rocek:1978nb} and is equivalent to the VA non-linearly realized theory in eq. \rf{VA}. 
But the setup studied by Dall'Agata {\it et al} in \cite{DallAgata:2022abm} is different from \cite{Komargodski:2009rz} since they promote the Lagrange multiplier superfield $T$ to a propagating field, following  \cite{Jaeckel:2002rm}. They add new terms in the \K potential 
\be
\Delta K = \beta \, T\bar T + g X\bar X T\bar T + {q\over 4} X\bar X X\bar X  \ .
\label{add}\ee
The first two terms make $T$ not a Lagrange multiplier anymore but a propagating superfield. Next the authors of \cite{DallAgata:2022abm} have changed variables to make $T$ a canonical superfield and computed the masses of all scalars. However, unlike the authors of the ERG approach \cite{Wetterich:1992yh}, 
Dall'Agata {\it et al} did not study the relation of their model to the original theory where $T$ is not a propagating field,  by looking at the limit 
\be
Z_T= \beta  \to 0 \ .
\label{4f}\ee 
Their model replacing the one in eq. \rf{SVA-KSxt} is 
\bea
{\cal S}_{Dall}^I &=& \int d^4  x\ls\int d^4 \theta  \ls \alpha  X  {\bar X} +   \beta  T  {\bar T} + g X\bar X T\bar T + {q\over 4}( X\bar X)^2\rs \right.\cr
&&\qquad\quad \ \left.+  \lp\int d^2 \theta \lp fX + {1\over 2} TX^2\rp + c.c\rp\rs \ .
\label{SFar0}\eea
The original action in eq. \rf{SVA-KSxt} is recovered for $\alpha=1, \beta=0, g=0, q=0$
\be
{\cal S}_{Dall}^I  (\alpha=1, \beta=0, g=0, q=0)=  {\cal S} (X, T) \ .
\label{DallVA}\ee 
The  ERG equations are imposed  \footnote{There is no underlying computation of the 1-loop quantum corrections of the kind we have explained in the example of the O'Raifeartaigh model. } in \cite{DallAgata:2022abm} and solved for space-time constants depending on the `renormalization group time' $t= \log {\Lambda\over \mu}$ so that the solutions for $\alpha, \beta, g, q, f$ are given as functions of $t$, $\mu^2$ and $\Lambda$. At $t=0$ the couplings take values $\alpha=1, \beta=0, g=0, q=0$.
 
To compute the masses of the 4 real scalars, which are present in the theory with 2 unconstrained chiral superfields, one changes variables so that all kinetic terms of new scalars in $\hat X, \hat T$ are canonical 
\bea
{\cal S}_{Dall}^{II} &=& \int d^4 \hat x\ls\int d^4\hat \theta  \ls  \hat X \hat {\bar X} +   \hat T \hat {\bar T} + \tilde \gamma \hat T \hat {\bar T}  \hat X \hat {\bar X}+  \tilde \zeta \hat X \hat {\bar X}  \hat X \hat {\bar X} \rs \right.\cr
&&\qquad\quad \ \left.+  \lp \int d^2 \hat \theta \lp \tilde f \hat X + \tilde g  \hat T \hat X^2 \rp + c.c\rp\rs \ .
\label{S2Far0}\eea
Here the relation between the rescaled  superfields and the rescaled superspace coordinates to the ones in eq. \rf{SFar0} is given by
 \be
 X= \mu{\hat X\over \sqrt\alpha}\, , \qquad T= \mu{\hat T\over \sqrt\beta}\, , \qquad  x= \hat x \mu^{-1} \, , \qquad  \theta= \hat \theta \mu ^{1/2} \ .
\label{hat} \ee
The new couplings in eq. \rf{S2Far0}  are
 \be
 \tilde \zeta = {\mu^2 q(\mu)\over 4\alpha^2} \, , \qquad \tilde \gamma = {\mu^2 g (\mu) \over \alpha \beta}\, , \qquad \tilde f = {e^{2t}  f\over \Lambda^2 \sqrt\alpha}\, , \qquad\tilde g= {1\over 2\alpha \sqrt \beta} \ ,
\label{par1} \ee 
where $\alpha, \beta, g, q, f$ are functions of $t$. Using the scalar potential $\hat V= \hat g^{i\bar \jmath} \partial_i \hat W \partial_{\bar \jmath}  \hat{ \bar W}$  based on eq. \rf{S2Far0} the masses of the two complex scalars in the $\hat X, \hat T$ superfields can be computed in the globally supersymmetric case. To find these at small $t$ we need to present the dependence of the couplings on $t$ as given by the solutions of the ERG equations in \cite{DallAgata:2022abm}
\be
\tilde \zeta = {1- e^{-2t}\over 4\alpha^2} \, , \qquad \tilde \gamma = {1- e^{-2t} \over \alpha \beta}\, , \qquad \tilde f = {e^{2t}  f\over  \Lambda^2\sqrt\alpha}\, , \qquad\tilde g= {1\over 2\alpha \sqrt \beta} \ ,
\label{par} \ee 
where 
\be
\alpha = 1- {1\over 16 \pi^2} + {1\over 8 \pi^2} \Big (t+ {1\over 2} e^{-2t}\Big)\, , \qquad \beta =  -{1\over 32 \pi^2} + {1\over 16 \pi^2} \Big (t+ {1\over 2} e^{-2t}\Big) \ .
\ee
At small $t$ one has
\be
\alpha \approx  1\, ,\qquad \beta \approx {1\over 16 \pi^2}  t^2 \ .
\label{limb}\ee
This leads to
\be
 \tilde \zeta \approx  {t\over 2} \, ,\qquad  \tilde \gamma \approx  {32 \pi^2\over   t} \, ,\qquad \tilde f \approx   {f \over \Lambda^2 }\, , \qquad\tilde g = {2\pi\over   t} \ .
\ee
Note that $\Lambda \geq \mu$. Therefore, $t = \log {\Lambda \over \mu}\geq 0$ and all $t$ dependent constants are positive.

The action in \rf {SFar0} at $t\neq 0$ is related to the action in \rf{S2Far0} by a change of variables in eq. \rf{hat} which is singular at $t=0$. The standard analysis of the supersymmetric theory proceeds using  \rf{S2Far0} with canonical kinetic terms. We  have shown in Sec. \ref{sec4} that the limit of this theory to $t\to 0$ is discontinuous and therefore there is no relation between the theory in \cite{DallAgata:2022abm} and the theory with a non-linearly realized supersymmetry.

\bibliographystyle{JHEP}
\bibliography{lindekalloshrefs}
\end{document}